\documentclass[12pt]{article}
\usepackage{amssymb,amsmath}
\usepackage{cite}
\usepackage{verbatim}
\usepackage{graphicx}

\usepackage{bm}

\let\OLDthebibliography\thebibliography
\renewcommand\thebibliography[1]{
	\OLDthebibliography{#1}
	\setlength{\parskip}{0pt}
	\setlength{\itemsep}{0pt plus 0.3ex}
}
  
\newcommand{\eq}{Eq.\,(\ref}

\newcommand{\dd}{\mbox{\rm d}}
\newcommand{\wg}{\wedge}
\newcommand{\gam}{\gamma}

\newcommand{\sg}{\sigma}
\newcommand{\Sg}{\Sigma}
\newcommand{\al}{\alpha}

\newcommand{\dg}{\dagger}

\newcommand{\tl}{\tilde}
\newcommand{\ul}{\underline}

\newcommand{\DD}{\mbox{\rm D}}

\newcommand{\p}{\partial}

\newcommand{\be}{\begin{equation}}
\newcommand{\bear}{\begin{eqnarray}}
\newcommand{\ear}{\end{eqnarray}}
\newcommand{\ee}{\end{equation}}
\newcommand{\lbl}{\label}
\newcommand{\bi}{\bibitem}
\newcommand{\ci}{\cite}
\newcommand{\pav}{Pav\v si\v c}
\newcommand{\vs}{\vspace}
\newcommand{\hs}{\hspace}

\newcommand{\bp}{{\bm p}}
\newcommand{\bP}{{\bm P}}
\newcommand{\hbp}{{\hat{\bm p}}}
\newcommand{\bx}{{\bm x}}
\newcommand{\barx}{{\bar x}}

\newcommand{\vphi}{\varphi}
\newcommand{\om}{\omega}
\newcommand{\Om}{\Omega}
\newcommand{\vac}{|0 \rangle}
\newcommand{\vacc}{\langle 0|}

\newcommand{\bbi}{{\bar i}}
\newcommand{\bpsi}{{\bar \psi}}
\newcommand{\hpsi}{{\hat \psi}}
\newcommand{\bbs}{{\bar *}}
\newcommand{\sqr}{\frac{1}{\sqrt{2}}}

\newcommand{\half}{\frac{1}{2}}

\begin{document}

\

\baselineskip .7cm 

\vs{8mm}

\begin{center}

{\LARGE A New Perspective on Quantum Field Theory Revelling Possible Existence of Another Kind of
	Fermions Forming Dark Matter}
	
\vs{3mm}

Matej Pav\v si\v c

Jo\v zef Stefan Institute, Jamova 39,
1000 Ljubljana, Slovenia

e-mail: matej.pavsic@ijs.si

\vs{6mm}

{\bf Abstract}
\end{center}

\baselineskip .5cm 

{\footnotesize 
Quantum fields are considered as generators of infinite-dimensional Clifford algebra $Cl(\infty)$, which
can be either orthogonal (in case of fermions) or symplectic (in case of bosons). A generic quantum
state can be expressed as a superposition of the basis elements of $Cl(\infty)$, the superposition
coefficients being multiparticle complex-valued wave functions. The basis elements, that are products of the generators of $Cl(\infty)$ in the Witt basis, act as creation
and annihilation operators. They
create positive and negative energy states that include the bare and the Dirac
vacuum as special cases. It is shown that the nonvanishing
electric charge arises from an extra dimension or from doubling the number of creation and annihilation operators, which brings
an extra imaginary unit $\bbi$ into the description. A further extension is to consider the $\bbi$ as
one of the quaternionic imaginary units and consider a generic state as having values in the quaternionic
algebra or, equivalently, in the complexified 2-dimensional Clifford algebra, $Cl(2)\otimes C$.
It contains two distinct fundamental representations of $SU(2)$, one associated with the weak isospin
doublet $(\nu_e,e^-)$, and the other one with the doublet of new leptons, denoted $(\epsilon^+,\nu_\epsilon)$,
that together with the new quarks $(u',d')$ can be identified with dark matter.
}

\vs{2mm}

{\it Keywords}:  Quantum field theory; orthogonal and symplectic Clifford algebras; negative energies; probability density versus charge density; new kind of fermions; dark matter.

Mathematical Subject Classification: 81T05; 81T08; 81S08; 81S10 

\baselineskip .6cm

\section{Introduction}

Foundations of quantum theory are still being challenged. In the last years, for instance, there has been a renewed discussion about whether wavefunction is necessarily complex, or perhaps it could be
real\ci{McKague,Renou}. That instead of a complex wave function, one
can consider twice as many real functions had already proposed in Refs.\ci{Stueckelberg}. Such a possibility
was ruled out by recent experiments\ci{Chen,Li}. On the other hand, the concept of the
relativistic wave function has been usually considered problematic or, in the case of the Dirac equation,
of limited validity, and asserted that the problems were resolved within
the framework of relativistic quantum field theory (see, e.g.,\ci{Itzykson}).
Because on the one hand the talk is about wave function (or its density matrix equivalent),
complex versus real-valued, and on the other hand the concept of wave function has no commonly accepted clear meaning in
relativistic quantum field theory, a further clarification of this issue is necessary. Despite the vast
literature on this topics\ci{NewtonWigner,Wightman,Kalnay,Ruijgrok,Barat,Mir-Kasimov,Cirilo-Lombardo,Herrmann,Fleming1,Fleming2,Malament,Monahan},
no general consensus has been established. An important insight can be
found in Ref.\ci{RosensteinHorwitz}, where the relation between the 
Newton-Wigner-Foldy wave function\ci{FoldyWaveFun}
and the Klein-Gordon wave function was considered. Another insight is provided in Ref.\ci{Fleming1,Fleming2}, where it
was realized that relativistic wave function and particle localization are defined with respect to a 3D surface
in spacetime. The meaning of localization and propagation of wave packets was investigated
in Refs.\ci{Karpov,Wagner, Eckstein, Al-Hashimi}.

The relation between relativistic wave function and quantum field theory
has been studied in Refs.\ci{PavsicRelatWaveFun1,PavsicRelatWaveFun2,PavsicRelatWaveFun3,PavsicStumbling}.
One of the key points is the realization that a {\it real} scalar field, $\vphi$, satisfying the Klein-Gordon
equation, is equivalent to a {\it complex} wave function $\psi$, satisfying the relativistic Schr\"odinger
equation\ci{Deriglazov}. Therefore, a two component scalar field, written as a complex field, cannot be identified with
a wave function---which is a right conclusion in the literature. But this does not mean that a (consistent)
relativistic wave function does not exist. Relativistic
wave function was identified\ci{PavsicRelatWaveFun3,PavsicStumbling} as a
superposition of $\vphi$ and $\Pi={\dot \vphi}$, namely, $\psi= \sqr (\sqrt{\om_\bx} \vphi + 
\frac{i}{\sqrt{\om_\bx}} \Pi)$, where $\om_\bx = \sqrt{m^2- \nabla^2}$. So constructed $\psi$ contains
only positive frequencies, so that the energy of such a state can only be positive. Upon quantization,
$\psi$ and $\psi^*$ become creation and annihilation operators, $a^\dg(\bx)$, $a(\bx)$, introduced in this form by
Jackiw\ci{Jackiw}, that create/annihilate a particle at position $\bx$. A single
particle state is\ci{PavsicRelatWaveFun1,PavsicRelatWaveFun2,PavsicRelatWaveFun3,PavsicStumbling}
$|\Psi \rangle = \int \dd^3 \bx \, \psi(t,\bx) a^\dg (\bx)\vac$. It satisfies the Schr\"odinger equation
$i \dd |\Psi \rangle/\dd t$ $= {\hat H} |\Psi \rangle$, where
${\hat H} = \half \int \dd^3 \bx \, ( a^\dg (\bx) \om_\bx a (\bx)$ $+ a(\bx) \om_\bx a^\dg (\bx) )$,
from which it follows that the wave function satisfies the equation $i \p \psi/\p t = \om_\bx \psi$.

In this paper we consider a 4-component {\it real} scalar field, $\vphi_\al$, $\al=1,2,3,4$, satisfying
the Klein-Gordon equation, and construct from $\vphi_\al$ a complex 4-component wave function
$\psi_\al =$  $\sqr \left (\sqrt{h} \vphi_\al + \frac{i}{\sqrt{h}} \Pi_\al \right )$ that satisifes
the Dirac equation $i \p \psi/\p t$ $= h \psi$, where $h= \beta m - i \al^i \p_i$, $i=1,2,3$.
We show that $\psi_\al$ so constructed is equal to the usual Dirac wave function.
Then instead of $h$ we take $h'= \beta \sqrt{m^2 + \p_i \p^i}$ and consider the wave function
$\psi'_\al =$  $\sqr \left (\sqrt{h'} \vphi_\al + \frac{i}{\sqrt{h'}} \Pi_\al \right )$.
The relation between $\psi'_\al$ and $\psi_\al$ is the Foldy-Wouthuysen transformation\ci{FWtransformation}, which
confirms the consistency of the procedure of constructing the Dirac wave function from a real 4-component
Klein-Gordon field. So constructed $\psi_\al$ contains positive and negative energies.
In Refs.\ci{Pagani,Smilga1,Smilga2,Smilga3,Smilga4,PavsicFirence,PavsicStable,PavsicPUReview,Kaparulin,Deffayet}
 it has been shown that negative energies do not lead
to instabilities in the presence of physically
realistic interaction potentials, bounded from below and from above. This opens the possibility to interprete
$\psi \equiv \psi_\al$ as probability density of finding a particle at position $\bx$ without the usual
limitations (see e.g., Ref.\ci{Itzykson}). It is
the zero component of the probability current density $j^\mu = {\bar \psi} \gam^\mu \psi$,
${\bar \psi} = \psi^\dg \beta$, $ \beta = \gam^0$. We then investigate the relation between the
probability density and charge density. A charge can be brought into the description if we consider 
a doublet of the Dirac wave functions, $\psi_n$, $n=1,2$. By the Noether procedure we can then derive
two conserved currents, namely, the probability current density, $j_n^\mu = {\bar \psi}_n \gam^\mu \psi_n$,
associated with the phase transfromations $\psi'_n = {\rm e}^{i \al} \psi_n$, $n=1,2$, of the group
$U_i (1)$, and the charge current density, associated with the rotations between $\psi_1$ and $\psi_2$ of the
group $U_\bbi$, where $\bbi$ is another imaginary unit entering the expression $\psi= \psi_1+ \bbi \psi_2$.
Alternatively, the electric charge comes from the 5th dimension. Then the charge current density
in four dimensions is
$j{_e n}^\mu =$ $e {\bar \psi_n} \gam^\mu \psi_n = e j_n^\mu$, where $e=p_5$, i.e., proportional to the
probability current density.

Next, we bring into the description the phase space basis vectors that can be either symplectic or
orthogonal\ci{Crumeyrole,PavsicOrthoSymp}. Their Witt basis equivalents are bosonic or fermionic creation and annihilation operators.
So we arrive at quantum field theory. We then concentrate the investigation on fermionic states
and their evolution equation. Next, we consider multi particle states as elements of an infinite
dimensional Clifford algebra $Cl(\infty)$. One such state is the product of infinitely many annihilation
operators, $\vac = \prod_{\bx,\al} a_\al (\bx)$, satisfying $a_\al (\bx \vac = 0.$.
In momentum representation we have $\vac_B = \prod_{\bp,\sg} b_\sg(\bp) d_{\sg}^\dg (\bp)$, $\sg=1,2$,
where $b_\sg (\bp)$ and $d_{\sg}^\dg (\bp)$ are the usual operators entering the expansion
of a Dirac field ${\hat \psi} (t,\bx)$. They satisfy $b_\sg (\bp) \vac_B = 0$, $d_{\sg}^\dg (\bp) \vac_B=0$,
where $\vac_B$ is the {\it bare vacuum} (see\ci{Hatfield}). The Fock space basis states are given in terms
of the products of $b_{\sg}^\dg (\bp)$ and $d_\sg (\bp)$, acting on $\vac_B$. One such state can be
$\prod_{\bp,\sg} d_\sg (\bp) \vac_B = \vac_D$, the so called Dirac vacuum, a sea of negative energy
particles. It is an idealization, because a realistic state is spread by a wave packet profile, which
determines an effective cutoff (see Fig.  ). For a single Dirac field, all those particle states and the antiparticle states
(``holes'' in the Dirac vacuum) have zero electric charge. Non zero electric charge arises if we
double the operators and the associated wave packet profiles, or if we consider an extra spacetime
dimension. In both cases a particle can have positive or negative charge, and its antiparticle
the opposite charge.

We then consider the presence of the electromagnetic field as arising from non local transformations
of the group $U_\bbi (1)$ acting within the doublet of the Dirac fields, written in the form
${\hat \psi} = {\hat \psi}_1 + \bbi {\hat \psi}_2$.

The next step is the realization that the $\bbi$ can be considered as one of three quaternionic imaginary
units $\bm i$, $\bm j$, $\bm k$. The wave function is then an element of quaternionic algebra. After
a suitable redefinition of the basis elements and the components, the wave function can be written
as an element of the complexified Clifford algebra $Cl(2)\otimes \mathbb{C}$. Within such a framework
one obtains two distinct spaces, associated with the first and the second left ideals of
$Cl(2)\otimes \mathbb{C}$, that form two distinct representations of $SU(2)$. Besides the usual
weak isospin lepton doublet $(\nu_e,e^-)$, there exists an additional lepton doublet $(\epsilon^+,\nu_\epsilon)$.
Analogously, besides the quark doublet $(u,d)$, this model incorporates an additional quark doublet
$(u',d')$. So this model predicts a new kind of fermions that could be invisible to the usual fermions.
The electric charge operator is the sum of ${\hat p}_5 = - i \p/\p x^5$ (identified with the hypercharge
operator ${\hat Y}/2$) and $I_3$ (one of the generators of $Cl(2)\otimes \mathbb{C}$). This scheme suggests
that also the weak interaction charges contain, besides the terms $g I_a$, $a=1,2,3$, also additional
terms due to the presence of extra dimensions. This is in agreement with the findings
of Refs.\ci{Cotaescu,PavsicKaluzaLong}.

In Sec.\,2 there is a discussion about the relation between the real Klein-Gordon field and the
complex wave function, which in Sec.\,3 is extended to the case of the Dirac equation. Then it is shown
how electric charge arises either from a doubling of the Dirac wave function or from an extra dimension.
A novel view of how quantum field theory enters the game is explained in Sec.\,4. The inclusion of
the electromagnetic field is considered in Sec.\,5. In Sec.\,6.1 it is shown that such an electromagnetic
field is only a part of the wider scheme which contains weak interaction. In Secs.\,6.2 and 6.3,
clarification is provided as to why negative energies are not problematic and why the concept of
relativistic wave packet localization, their propagation and Lorentz covariance make sense. Finally, it is
pointed out that the leakage of probability outside the light cone does not imply a breakdown of
causality on a macroscopic level.

 \section{Complex versus real wave function}
 
 In literature on quantum mechanics, wave function is usually taken to be complex. It satisfies the Schr\"odinger
 equation
 \be
   i \frac{\p \psi}{\p t} = h \psi,
\lbl{2.1}
\ee
where $h$ is a Hamilton operator. Writing the wave function $\psi$ in terms of its real and imaginary
part,
\be
  \psi = \psi_R + i \psi_I ,
\lbl{2.2}
\ee
the equation (\ref{2.1}) can be written as the following system of two equations:
\be
  \frac{\p \psi_R}{\p t} = h \psi_I ,
\lbl{2.4}
\ee
\be
\frac{\p \psi_I}{\p t} = -h \psi_R .
\lbl{2.5}
\ee
Differentiating Eq.\.(\ref{2.4}) with respect to time $t$ and using Eq.\,(\ref{2.5}), we obtain
\be
  \frac{\p^2 \psi_R}{\p t^2} + h^2 \psi_R = 0 .
\lbl{2.6}
\ee
In the above derivation we have assumed $\p h/\p t =0$.

Alternatively, from Eq.\,(\ref{2.4}) we express $\psi_I$ in terms of $\psi_R$ according to
\be
  \psi_I = h^{-1} {\dot \psi}_R ,
\lbl{2.6a}
\ee
and plug it into Eq.\,(\ref{2.5}). So we obtain
\be
  h^{-1} {\ddot \psi}_R + h \psi_R = 0 .
\lbl{2.7}
\ee
Multiplying the latter equation from the left by $h$ we obtain Eq.\,(\ref{2.6}).

One complex equation (\ref{2.1}) or, equivalently, the system of two first-order differential
equations with respect to time, can be replaced by one second-order equation (\ref{2.6})
for a real variable $\psi_R$. Using the analogous procedure for the imaginary part, we find
that it satisfies
\be
\frac{\p^2 \psi_I}{\p t^2} + h^2 \psi_I = 0 .
\lbl{2.8a}
\ee

The Hamilton operator of nonrelativistic quantum mechanics is
\be
  h= \frac{\hbp^2}{2 m} + V(\bx) ,
\lbl{2.9}
\ee
where $\hbp = - i \nabla$. For such Hamilton operator it make much more sense to use
the first order Schr\"odinger equation (\ref{2.1}) with a {\it complex wave function}
than to use the second order equation (\ref{2.6} with a real function $\psi_R$, because
$h^2$ then containes the terms with $\nabla^4$.

Taking $V=0$, the $h$ of Eq.\,(\ref{2.9}) is obtained as an approximation from the expression of
the relativistic Hamilton operator
\be
  h = \sqrt{\hbp^2 + m^2} = m \left ( 1 + \frac{\hbp^2}{2 m} + ... \right ) .
\lbl{2.9a}
\ee
Its square is $h^2 = \hbp^2 + m^2$. Equation (\ref{2.6}) then becomes the Klein-Gordon
equation for a {\it real} field $\psi_R$:
\be
  \frac{\p^2 \psi_R}{\p t^2} - \nabla^2 \psi_R + m^2 \psi_R = 0.
\lbl{2.10}
\ee

We have seen that a complex-valued wave function $\psi$ that satisfies the Schr\"odinger
equation (\ref{2.1}) can be written in terms of a real field $\psi_R$ and its time
derivative ${\dot \psi}_R$ as a superposition of the form
\be
   \psi = \psi_R + i h^{-1} {\dot \psi}_R .
\lbl{2.11}
\ee
The field $\psi_R$ satisfies the second prder equation (\ref{2.6}), which in the case of
$h= \sqrt{m^2 - \nabla^2}$ is the Klein-Gordon equation (\ref{2.10}).

The opposite procedure is to start with the second order equation (\ref{2.6}) and use
the variables
\be
  \psi_R~,~~~~~~~~\psi_I = h^{-1}{\dot \psi}_R ,
\lbl{2.12}
\ee
in term of which the second order equation (\ref{2.10}), i.e., ${\ddot \psi}_R + h^2 \psi_R = 0$,
becomes the system of two first order equations (\ref{2.4}),(\ref{2.5}) that is equjivalent to
the Schr\"odinger equation (\ref{2.1}).

Let us now introduce a field $\vphi$ according to the following functional transformation:
\be
  \psi_R = \sqrt{\frac{h}{2}} \vphi~,~~~~~~~~\psi_I = \frac{1}{\sqrt{2 h}} {\dot \vphi} ,
\lbl{2.12a}
\ee
The function $\psi_R$ satisfies the Klein-Gordon equation, and so does also $\vphi$.
\be
  {\ddot \vphi} + h^2 \vphi = 0.
 \lbl{2.12b}
 \ee
We will assume that $\vphi$ is a scalar under Lorentz transformations. According to the relations
(\ref{2.11}) and (\ref{2.12a}), $\psi_R$ is then not a scalar field.

The action for the scalar field is
\be
  I[\vphi] = \frac{1}{2} \int \dd t\, \dd^3 \bx \left ( {\dot \vphi}^2  - \vphi\, \om_\bx^2 \vphi \right ),
\lbl{2.13}
\ee
where $h=\om_\bx \equiv \sqrt{m^2 - \nabla^2}$. Introducing the canonically conjugate momentum
$\Pi= \p {\cal L}/\p {\dot \vphi} = {\dot \vphi}$, we can write the {\it  phase space} form of the action
(\ref{2.13}):
$$
  I[\vphi,\Pi]=  \int \dd t \, \dd^3 \bx \left [ \Pi {\dot \vphi} 
  - \frac{1}{2} (\Pi^2 + \vphi \om_\bx^2 \vphi ) \right ]$$
\be
  \hs{1cm}= \int \dd t \, \dd^3 \bx \left [ \frac{1}{2} (\Pi {\dot \vphi} - \vphi {\dot \Pi})
  - \frac{1}{2} \left ( \Pi^2 + (\om_\bx \vphi) (\om_\bx \vphi) \right ) \right ] .
\lbl{2.14}
\ee
In the last step, we have performed the integration per partes and omitted the surface terms.
For that purpose we expanded $\om_\bx \equiv h$ according to (\ref{2.9a}).

The action (\ref{2.13}) or (\ref{2.14}) has the form of the action for an uncountably infinite
set of harmonic oscillators, one for each $\bx$, and having the frequency $\om_\bx$. 

Introducing the new variables\ci{Jackiw}
\be
  \psi = \frac{1}{2} \left ( \om_\bx^{1/2} \vphi + i \om_\bx^{-1/2} \Pi \right ) ,
\lbl{2.15}
\ee
\be
   \psi^* = \frac{1}{2} \left ( \om_\bx^{1/2} \vphi - i \om_\bx^{-1/2} \Pi \right ) ,
\lbl{2.16}
\ee 
the phase space action (\ref{2.14}) becomes
\be
  I[\psi,\psi^*] = \int \dd t \, \dd^3 \bx \left [ \frac{i}{2} (\psi^* {\dot \psi}
   - {\dot \psi}^* \psi) - \psi^* \om_\bx \psi \right ].
\lbl{2.17}
\ee
Variation of the latter action with respect to $\psi$ and $\psi^*$, respectively, gives the following
equations of motion:
\be
   i {\dot \psi} = \om_\bx \psi ,
\lbl{2.18a}
\ee
\be
  - i {\dot \psi}^* = \om_\bx \psi^* .
\lbl{2.18b}
\ee
By plugging Eqs.\,(\ref{2.15}),(\ref{2.16}) into Eqs.\,(\ref{2.18a}), (\ref{2.18b})
we obtain
\be
  {\dot \vphi} = \Pi ,
\lbl{2.19}
\ee
\be
  {\dot \Pi} = - \om_\bx^2 \vphi ,
\lbl{2.20}
\ee
which also follows directly from the action $I[\vphi,\Pi]$  (Eq.\,(\ref{2.14})).

The Hamiltonian belonging to the action (\ref{2.14}) is
\be
   H[\vphi,\Pi] = \int \dd^3 \bx\, \frac{1}{2} \left ( \Pi^2 + \vphi \om_\bx^2 \vphi \right ).
\lbl{2.21}
\ee
Written in terms of the variables $\psi$, $\psi^*$ the Hamiltonian becomes
\be
  H[\psi,\psi^*] = \int \dd^3 \bx\, \psi^* \, \om_\bx \psi .
\lbl{2.22}
\ee
The new variable $\psi$ defined by Eq.\,(\ref{2.15}) is the same as the wave function $\psi$ defiined in Eq.\,(\ref{2.2}), which in view of (\ref{2.12a}) can be written as
\be
  \psi = \frac{1}{\sqrt{2}} ( h^{1/2} \vphi + i h^{-1/2} {\dot \vphi} ),
\lbl{2.25a}
\ee
where in particular it could be $h= \om_\bx$. Thus, Eqs.\,(\ref{2.18a}) and (\ref{2.18b}) are, respectively,
the Shr\"odinger equation for the wave function $\psi$ and its complex conjugate $\psi^*$.

That the Klein-Gordon equation can be written as a system of two first-order equation is known
in the literature\ci{FoldyWaveFun}. The usual procedure goes as follows. One starts from the Klein-Gordon
equation for a {\it complex} field $\vphi$,
\be
  {\ddot \vphi} + h^2 \vphi = 0 ,
\lbl{2.24}
\ee
where $h= \sqrt{m^2 - \nabla^2}$, and introduces the new variables
\be
  \chi_1 = \left (i \frac{\p}{\p t} + h \right ) \vphi,
\lbl{2.25}
\ee
\be
\chi_2 = \left (-i \frac{\p}{\p t} + h \right ) \vphi,
\lbl{2.26}
\ee
in term of which Eq.\,(\ref{2.24}) can be rewritten as
\be
   \left (-i \frac{\p}{\p t} + h \right ) \chi_1 = 0 ,
\lbl{2.27}
\ee
\be
\left (i \frac{\p}{\p t} + h \right ) \chi_2 = 0 .
\lbl{2.28}
\ee
In matrix form this becomes\ci{Feshbach,FoldyWaveFun}
\be
  i\frac{\p}{\p t}     \begin{pmatrix}
                     	\chi_1\\
                    	\chi_2
                         \end{pmatrix}
   = h  \begin{pmatrix}
   	     1 & 0\\
   	     0 & -1
   \end{pmatrix}
   \begin{pmatrix}
   \chi_1\\
   \chi_2
\end{pmatrix} .
\lbl{2.29}
\ee

What is not considered in the literature is the fact that for a {\it real} Klein-Gordon field,
$\vphi = \vphi^*$, the field $\chi_1$ is the complex conjugate of the field $\chi_2$, i.e.,
\be
  \chi_1 = \chi_2 \equiv \chi .
\lbl{2.30}
\ee
Therefore the second equation (\ref{2.28}) is the complex conjugate of the equation (\ref{2.27}),
and not an independent equation. If $\chi_1$ as a solution of Eq.\,(\ref{2.27}) is known, then
also $\chi_2 =\chi_1^*$ is known. Thus only one of the two equations (\ref{2.27}),(\ref{2.28}) is
sufficient to replace the Klein-Gordon equation if the field is real.

If $\vphi$ is {\it real}, then the function $\chi_1 \equiv \chi$ is proportional to the
wave function $\psi$ given in Eq.\,(\ref{2.15}), the proportionality factor being $\sqrt{2 \om_\bx}$, so
that $\chi = \sqrt{2 \om_\bx} \psi$.

Using the general solution of the Klein-Gordon equation for a real field,
\be \vphi = \int \frac{\dd^4 p}{(2 \pi)^4} \delta (p^2 - m^2) {\rm e}^{-i px} c (p)
    = \int \frac{\dd^3 \bp}{(2 \pi)^4} \frac{1}{2 \om_\bp}
   \left ( c(\om_\bp,\bp) {\rm e}^{-i px} + c(-\om_\bp,\bp) {\rm e}^{i px} \right ) ,
\lbl{2.31}
\ee
we find
\bear
  &&\psi (t,\bx) = \sqrt{\frac{\om_\bx}{2}} \left (\vphi + i \om_\bx^{-1} {\dot \vphi} \right )
       = 2 \sqrt{\frac{\om_\bx}{2}} \int \frac{\dd^3 \bp}{(2 \pi)^4} \frac{1}{2 \om_\bp}
       c(\om_\bp,\bp) {\rm e}^{-i px}
       = \sqrt{2\om_\bx} \vphi^{(+)}, 
\lbl{2.32}\\
    &&\psi^* (t,\bx) =  \sqrt{\frac{\om_\bx}{2}} \left (\vphi -i \om_\bx^{-1} {\dot \vphi} \right )
    = 2 \sqrt{\frac{\om_\bx}{2}} \int \frac{\dd^3 \bp}{(2 \pi)^4} \frac{1}{2 \om_\bp}
       c(-\om_\bp,\bp) {\rm e}^{i px}    =\sqrt{2\om_\bx} \vphi^{(-)},
\lbl{2.32a}
\ear
where $\vphi^{(+)}$ is the positive frequency part of the solution (\ref{2.31}), and
$\vphi^{(-)}$ the negative part.
Rewriting $\vphi^{(+)}$ according to
\be
  \vphi^{(+)}= \int \frac{\dd^3 \bp}{(2 \pi)^4} \frac{1}{2 \om_\bp}
  c(\om_\bp,\bp) {\rm e}^{-i px} =
  \int \frac{\dd^3 \bp}{\sqrt{(2 \pi)^3 2 \om_\bp}} a(\bp) {\rm e}^{-i px} ,
\lbl{2.33}
\ee
where
    $a (\bp) = \frac{1}{2 \pi}\frac{c(\bp)}{\sqrt{(2 \pi)^3 2 \om_\bp}}$,
equations (\ref{2.32})--(\ref{2.33}) give
\be
   \psi (t,\bx) =  \int \frac{\dd^3 \bp}{\sqrt{(2 \pi)^3}} a(\bp) {\rm e}^{-i px} = a(t,\bx),
\lbl{2.33a}
\ee
\be
\psi^* (t,\bx) =  \int \frac{\dd^3 \bp}{\sqrt{(2 \pi)^3}} a^*(\bp) {\rm e}^{i px}= a^*(t,\bx)  .
\lbl{2.33ab}
\ee
We see that the wave function contains only positive energies (frequencies) $\om_\bp$.

{\it Nonrelativistic approximation}

Let us take the ansatz
\be
  \psi = {\rm e}^{-i m t} \phi,
\lbl{3.34a}
\ee
and insert it into the relativistic Shr\"odinger equation
\be
   i {\dot \psi} = \sqrt{m^2 - \nabla^2} \psi ,
\lbl{2.34}
\ee
which is just a rewritten equation (\ref{2.18a})\footnote{
   Recall that $\om_\bx = \sqrt{m^2 - \nabla^2}$.} which, as we have seen, is equivalent to the
Klein-Gordon equation for a {\it real} scalar field $\vphi$.
Then we obtain
\be
   i {\dot \phi} = \sqrt{m^2 - \nabla^2}\, \phi - m \phi .
\lbl{2.35}
\ee
Expanding
\be
   \sqrt{m^2 - \nabla^2} = m \left ( 1 - \frac{\nabla^2}{2 m} + ... \right ),
\lbl{3.36a}
\ee
Eq.\,(\ref{2.35}) becomes
\be
 i {\dot \phi} =\left (- \frac{\nabla^2}{2 m} + ... \right ) \phi
\lbl{2.37}
\ee
Neglecting the higher-order terms, the above equation is the nonrelativistic Schr\"odinger
equation.

In the literature, the transition from the Klein-Gordon equation to its nonrelativistic
approximation is not done according to the above procedure. Usually, it is started from a
{\it complex} field $\vphi$ satisfying the Klein-Gordon equation and taking the Ansatz
$\vphi= {\rm e}^{-i m t} \phi$, which leads to the equation
\be
   \frac{1}{2m} {\ddot \phi} - i {\dot \phi}  - \frac{1}{2m} \nabla^2 \phi =0 .
\lbl{2.38}
\ee
In the nonrelativistic approximation the first term can be neglected and the above
equation becomes the Schr\"odinger equation. 

\vs{2mm}

{\it A common mistake concerning the Klein-Gordon field}

We have seen that the real field $\vphi$, satisfying the second order equation (\ref{2.12b}),
is itself not the relativistic wave function. The relativistic wave function is the
superposition (\ref{2.15}) of $\vphi$ and its canonical momentum $\Pi={\dot \vphi}$.
The field $\vphi$ is real and the wave function
\be
  \psi = \sqrt{\frac{\om_\bx}{2}} \left ( \vphi + i \om_\bx^{-1} {\dot \vphi} \right )
\lbl{2.38a}
\ee
is complex. In the case of two real fields $\vphi_1$ and $\vphi_2$, we obtain two
complex wave functions
\be
\psi_1 = \sqrt{\frac{\om_\bx}{2}} \left ( \vphi_1 + i \om_\bx^{-1} {\dot \vphi_1} \right ) ~,
~~~~~~~~
\psi_2 = \sqrt{\frac{\om_\bx}{2}} \left ( \vphi_2 + i \om_\bx^{-1} {\dot \vphi_2} \right )
\lbl{2.40}
\ee

The fields $\vphi_1$ and $\vphi_2$ can be written together as a complex Klein-Gordon field
\be
  \vphi = \vphi_1 + {\bar i} \vphi_2 ,
\lbl{2.41}
\ee
where the imaginary unit $\bbi$, $\bbi^2 = -1$, in general is not the same one as the $i$
occurring in the wave functions $\psi_1$ and $\psi_2$, defined in Eq.\,(\ref{2.40}).
Each of those wave functions satisfies the relativistic Schr\"odinger equation (\ref{2.34})
which can be written as the system of two equations (\ref{2.19}),(\ref{2.20}) for each pair
of the real variable $\vphi_n$ and ${\dot \vphi}_n = \Pi_n$, n=1,2.
Altogether we have four real variables $\vphi_1$, $\Pi_1$, $\vphi_2$, $\Pi_2$.
The fact that we can combine the variables $\vphi_i$ and $\Pi$ into the complex variables
$\psi_n = \sqrt{\frac{\om_\bx}{2}} \left ( \vphi_n + i \om_\bx^{-1} {\dot \vphi_n} \right )$,
$n=1,2$, should not be confused with the fact that also $\vphi_1$ and $\vphi_2$ can be
combined into a complex variable \`a la (\ref{2.41}). In general, because there are four
independent real variables $\vphi_n$, ${\dot \vphi}_n = \Pi_n$, $n=1,2$, when writing them in terms of
complex numbers, one has to distinguish between the imaginary units $i$ and $\bbi$.

The above points are not properly addressed in the literature. To illustrate further the
usual misconceptions, let us act with the derivative $\p/\p t$ on the left and on the right hand
side of Eq.\,(\ref{2.34}). We obtain
\be
   {\ddot \psi} - \nabla^2 \psi + m^2 \psi = 0 .
\lbl{2.42}
\ee
This equation has the form of the Klein-Gordon equation for a complex field $\psi$.
However, the complex field $\psi$ is composed of the {\it real} fields $\vphi$ and ${\dot \vphi}$
according to (\ref{2.38a}). Assuming that $\vphi$ is a scalar field and transforms under the Lorentz transformations as $\vphi'(t',\bx') = \vphi (t,\bx)$, then by its construction, $\psi$ does
{\it not} transforms under a Lorentz transformation as a scalar field, i.e., not as
$\psi'(t',\bx') = \psi (t,\bx)$. It transforms in a more complicated way that depends on
the split of spacetime into a time $t$ and a space $\bx$, which in every Lorentz system is
different. We will discuss this in Sec.\,6.

If the scalar field $\vphi$ is a two-component object $\vphi = (\vphi_1,\vphi_2)$ and can
thus be written as a complex field $\vphi= \vphi_1 + \bbi \vphi_2$, it cannot be interpreted
as a relativistic wave function. The attempts to interpret a complex $\vphi$ as a wave function
have failed, which led people to the incorrect conclusion that a relativistic wave function makes no sense and that a consistent relativistic quantum mechanics does not exist. As a follow-up to the previous work\ci{PavsicRelatWaveFun1,PavsicRelatWaveFun2,PavsicRelatWaveFun3},
we provide in this paper further arguments that relativistic quantum mechanics makes sense and how
it arises within quantum field theory as the mechanics of wave packet profiles.

The generator of the transformation
\be
   \vphi \rightarrow {\rm e}^{\bbi \beta} \vphi ,
\lbl{2.43}
\ee
i.e., a rotation between $\vphi_1$ and $\vphi_2$ is proportional to the electric charge:
\be
  G_e = - \beta \int \dd^3 \bx \,\left ( {\dot \vphi}_1 \vphi_2 - {\dot \vphi_2} \vphi_1 \right ) =
     - i \beta \int \dd^3 \bx \,\left ( { \psi}_1^* \psi_2 - {\psi_2}^* \psi_1 \right ) .
\lbl{2.45}
\ee

The generator of the transformation
\be
   \psi_1 = {\rm e}^{i \alpha} \psi_1 ~,~~~~~~~\psi_2 = {\rm e}^{i \alpha} \psi_2 ,
\lbl{2.46}
\ee
i.e., a rotation between $\psi_{nR}$ and $\psi_{nI}$, $n = 1,2$, is proportional to the
conserved probability:
\be
  G = - i \alpha \int \dd^3 \bx \,\left ( { \psi}_1^* \psi_1 + {\psi_2}^* \psi_2 \right ) .
\lbl{2.47}
\ee
Here "*" performs the complex conjugation with respect to $i$ and with respect to $\bbi$.

In Refs.\ci{PavsicStumbling} it was shown explicitly how the generators $G_e$ and $G$ can
be derived from the Noether principle from the actions (\ref{2.13}) and (\ref{2.17}) in which
$\vphi$ and $\psi$ are considered as two component objects. Because the actions
(\ref{2.13}),(\ref{2.17}) are invariant unders the transformations (\ref{2.43}) and (\ref{2.46}),
the generators $G_e$ and $G$ are conserved.

Introducing the new variables
\be
  \chi_+ = \frac{\psi_1 - \bbi \psi_2}{\sqrt{2}}~,~~~~~
  \chi_- = \frac{\psi_1 + \bbi \psi_2}{\sqrt{2}} ,
\lbl{2.48}
\ee
we find
\be
  G_e = \beta \int \dd^3 \bx \, \left ( \chi_+^* \chi_+ - \chi_-^* \chi_- \right ),
\lbl{2.49}
\ee
and
\be
  G = \alpha \int \dd^3 \bx \, \left ( \chi_+^* \chi_+ + \chi_-^* \chi_- \right ).
\lbl{2.50}
\ee

Here $\chi_+^* \chi_+$ 
and $\chi_-^* \chi_-$ are the probability densities of finding,
respectively, a positively and negatively charged particle at the position $\bx$.
In the expression for $G_e$ there occurs the difference of those probabilities densities
which gives the net charge of the system.

\section {A new view on the Dirac equation}
\subsection{The Dirac equation from the 4-component Klein-Gordon equation}

We will now consider a particular choice of Hamilton operator, namely,
\be
  h= \beta m  - i \alpha^i \p_i ,
\lbl{3.1}
\ee
where $\alpha^i$, $i = 1,2,3$ and $\beta$ are the Dirac matrices. Let us start from the
Klein-Gordon equation
\be
  {\ddot \vphi} + h^2 \vphi = 0 ,
\lbl{3.2}
\ee
where $\vphi$ now denotes four {\it real} fields, $ \vphi = \vphi^\alpha$, $\alpha = 1,2,3,4$.
Following the preceding procedure, let us introduce the wave function
\be
  \psi = \sqrt{\frac{h}{2}} \left (\vphi + i h^{-1} {\dot \vphi} \right ) ,
\lbl{3.3}
\ee
which satisfies the Schr\"odinger equation
\be
  i {\dot \psi} = h \psi .
\lbl{3.4}
\ee
Using Eq.\,(\ref{3.3}), we have
\be
  i \left ({\dot \vphi} + i h^{-1} {\ddot \vphi} \right ) = h \left ( \vphi + i h^{-1} {\dot \vphi} ,
   \right )
\lbl{3.4a}
\ee
which gives the Klein-Gordon equation (\ref{3.2}). The first-order differential equation
(\ref{3.4}) is thus equivalent to the second order equation (\ref{3.2}). If $h$ is that of
Eq.\,(\ref{3.1}), then the $\vphi$ and hence the $\psi$ are 4-component objects on which
act the Dirac matrices $\alpha^i$ and $\beta$.

A general solution of the Klein-Gordon equation is
\be
  \vphi = \int \frac{\dd^3 \bp}{\sqrt{(2 \pi)^3 2 \om_\bp}}	\left ( a (\bp) {\rm e}^{-i p x}
  + a^* (\bp) {\rm e}^{i p x} \right ) .
\lbl{3.5}
\ee
Inserting the latter expression into Eq.\,(\ref{3.3}), we obtain the following wave
function: 
$$
   \psi = \int \frac{\dd^3 \bp}{\sqrt{(2 \pi)^3 2 \om_\bp}}  \left [
   \sqrt{\frac{\beta m - {\bm \alpha} \bp}{2}} \left ( 1 + \frac{\om_\bp}{\beta m - {\bm \alpha} \bp}
    \right ) a(\bp) {\rm e}^{-i p x}  \right .  \hs{2cm}$$
$$ \left .  \hs{3cm}  + \sqrt{\frac{\beta m + {\bm \alpha} \bp}{2}}\left ( 1 - \frac{\om_\bp}{\beta m + {\bm \alpha} \bp} \right ) a^*(\bp) {\rm e}^{i p x} \right ]$$
$$
\hs{.9cm}= \int \frac{\dd^3 \bp}{\sqrt{(2 \pi)^3 2 \om_\bp}} \frac{1}{\sqrt{2}} 
  \left [ \frac{1}{\sqrt{\beta m -{\bm \alpha} \bp}}
	(\beta m -{\bm \alpha} \bp + \om_\bp) a(\bp) {\rm e}^{-i p x}  \right . \hs{2cm}$$
\be
  \left . \hs{3cm}	+\frac{1}{\sqrt{\beta m +{\bm \alpha} \bp}}
	(\beta m +{\bm \alpha} \bp - \om_\bp) a^*(\bp) {\rm e}^{i p x} \right ] ,
\lbl{3.6}
\ee
where $ {\bm \alpha} \bp \equiv \gam^0 \gam^i p_i$.
This is a general solution of the first order equation (\ref{3.4}). It contains positive
and negative energies $p^0= \pm \om_\bp$. 

From the relation
\be
   (\beta m \mp {\bm \alpha \bp})^2 = m^2 + \bp^2 = \om_\bp^2 ,
\lbl{3.6a}
\ee we have
\be
  (\beta m - {\bm \alpha}\bp) (\beta m -{\bm \alpha} \bp + \om_\bp)
  = \om_\bp  (\beta m -{\bm \alpha} \bp + \om_\bp) ,
\lbl{3.6b}
\ee
\be
(\beta m + {\bm \alpha}\bp) (\beta m +{\bm \alpha} \bp - \om_\bp)
= -\om_\bp  (\beta m -{\bm \alpha} \bp + \om_\bp) ,
\lbl{3.6c}
\ee
and also
\be
(\beta m - {\bm \alpha}\bp)^{-1/2} (\beta m -{\bm \alpha} \bp + \om_\bp)
= {\om_\bp}^{-1/2}  (\beta m -{\bm \alpha} \bp + \om_\bp) ,
\lbl{3.6d}
\ee
\be
(\beta m + {\bm \alpha}\bp)^{-1/2} (\beta m +{\bm \alpha} \bp - \om_\bp)
= (-\om_\bp)^{-1/2}  (\beta m -{\bm \alpha} \bp + \om_\bp) ,
\lbl{3.6e}
\ee
Using the latter relations in Eq.\,(\ref{3.6a}), we obtain obtain the following
expression for the wave function:
$$
  \psi = \int \frac{\dd^3 \bp}{\sqrt{(2 \pi)^3}}\frac{1}{2 \om_\bp}
  	 \left [(\beta m -{\bm \alpha} \bp + \om_\bp) a(\bp) {\rm e}^{-i p x} +
  	 \frac{1}{i} (\beta m +{\bm \alpha} \bp - \om_\bp) a^*(\bp) {\rm e}^{i p x} \right ]
$$
\be
   = \int \frac{\dd^3 \bp}{\sqrt{(2 \pi)^3}}\frac{1}{2 \om_\bp}
   \left [(\gam^\mu p_\mu + m) \beta a(\bp) {\rm e}^{-i p x} + \frac{1}{i} (- \gam^\mu p_\mu + m) \beta a^* (\bp)
   {\rm e}^{i p x} \right ].
\lbl{3.6f}
\ee
Inserting the expression (\ref{3.6f}) for the $\psi$ into Eq.\,(\ref{3.4}) and using (\ref{3.6a})
we verify that it is the solution of the relativistic Schr\"odinger equation (\ref{3.4})
with $h = \beta m - i \alpha^i \p_i$. This means that Eq.\,(\ref{3.4}) is the Dirac equation,
and the $\psi$ a spinor wave function.

Let us now investigate how the wave function (\ref{3.6f}) is related to the usual expression for the
Dirac wave function
\be
  \psi = \int \frac{\dd^3 \bp}{(2 \pi)^3}\frac{m}{E}
  \left ( b_\sg (\bp) u^\sg (\bp) {\rm e}^{-i p x} + d^* (\bp) v^\sg (\bp) {\rm e}^{i p x} \right ),
  \lbl{3.17}
\ee
where $\sg=1,2$ and
\be
u^\sg (\bp) = \frac{\gam^\mu p_\mu + m}{\sqrt{2m (m+E)}} u^\sg (m,0) ,
\lbl{3.20}
\ee
\be
v^\sg (\bp) = \frac{-\gam^\mu p_\mu + m}{\sqrt{2m (m+E)}} v^\sg (m,0) ,
\lbl{3.21}
\ee
are, respectively, the positive and negative energy solutions of the Dirac equation, whilst
$b(\bp)$ and $d^* (\bp)$ are the corresponding wave packet profiles.

First let us consider a particular case of the wave packet, such that
\be
  a(\bp) = \delta^3 (\bp) a(0) ,
\lbl{3.22}
\ee
and insert it into the wave function (\ref{3.6f}). After integrating out $\bp$, we obtain
$$
  \psi = \frac{1}{\sqrt{(2 \pi)^3}} \frac{1}{2 m} \left [ (\beta + 1) m {\rm e}^{-i m t} a(0)
  + \frac{1}{i} (\beta - 1) m {\rm e}^{i m t} a^* (0) \right ]
$$
\be
  =  \frac{1}{\sqrt{(2 \pi)^3}} \left [ 
    \begin{pmatrix}
  	a_1(0) \\
  	a_2 (0) \\
  	0 \\
  	0 
  \end{pmatrix} {\rm e}^{-i m t}
 + \frac{-1}{i} 
  \begin{pmatrix}
 	0\\
 	0 \\
 	a_3^*(0) \\
 	a_4^*(0)
 \end{pmatrix}  {\rm e}^{i m t}  \right ] .
\lbl{3.23}
\ee
Comparing this with the Dirac spinors (\ref{3.20}),(\ref{3.21}) for $\bp=0$, we find the relations
\be
  \sqrt{(2 \pi)^3} a_1(0) = b_1 (0)~,~~~~~\sqrt{(2 \pi)^3} a_2 (0) = b_2 (0) ,
\lbl{3.24}
\ee
\be
   i \sqrt{(2 \pi)^3} a_3^*(0) = d_1^* (0)~,~~~~~i \sqrt{(2 \pi)^3} a_4^* (0) = d_2^* (0) .
\lbl{3.25}
\ee

In general, for $\bp\neq 0$, the terms in Eq.\,(\ref{3.6f}) give
\be
  (\gam^\mu p_\mu + m) \beta a(\bp) = 
   \begin{pmatrix}
  	m + \om_\bp && \sg^i p_i \\
  	-\sg^i p_i  && m - \om_\bp
  	\end{pmatrix}
  \beta	\begin{pmatrix}
  			{\bar a}(\bp)\\
  			{\ul a}(\bp) 
  		\end{pmatrix} ,
\lbl{3.26a}
\ee
\be
(-\gam^\mu p_\mu + m) \beta a(\bp) = 
\begin{pmatrix}
	m - \om_\bp && -\sg^i p_i \\
	\sg^i p_i && m + \om_\bp
\end{pmatrix}
\beta	\begin{pmatrix}
	{\bar a}^*(\bp)\\
	{\ul a}^*(\bp) 
\end{pmatrix} ,
\lbl{3.26b}
\ee
where ${\bar a} = \begin{pmatrix} a_1\\ a_2 \end{pmatrix} $ , ${\ul a} = \begin{pmatrix} a_3\\ a_4 \end{pmatrix} $,
${\bar a}^* = \begin{pmatrix} a_1^*\\ a_2^* \end{pmatrix} $, and ${\ul a}^* = \begin{pmatrix} a_3^*\\ a_4^* \end{pmatrix} $, which differ from the Dirac spinors.

Introducing new variables $A(\bp)$ and $B(\bp)$ according to
\be
  a(\bp) = \frac{\om_\bp + m + \gam^i p_i}{2 \om_\bp (\om_\bp + m )} A(\bp),
\lbl{3.27}
\ee
\be
a^*(\bp) = \frac{-\om_\bp + m - \gam^i p_i}{2 \om_\bp (\om_\bp + m )} B(\bp),
\lbl{3.28}
\ee
equation (\ref{3.6f}) becomes
$$
\psi = \int \frac{\dd^3 \bp}{\sqrt{(2 \pi)^3}}\frac{1}{\sqrt{2 \om_\bp (\om_\bp +m))}}
\left [(\gam^\mu p_\mu + m) \frac{1+\beta}{2} A(\bp) {\rm e}^{-i p x} \right .
$$
\be
  \hs{2cm} \left .   + \frac{1}{i} (- \gam^\mu \bp + m) \frac{1-\beta}{2} B(\bp)
{\rm e}^{i p x} \right ].
\lbl{3.29}
\ee
Let us now observe that
\be
   \frac{1+\beta}{2} A(\bp) =
   \begin{pmatrix} A_1 (\bp)\\ A_2 (\bp)\\ 0\\ 0 \end{pmatrix} = A_\sg (\bp) u^\sg (0) ~,
 ~~{\rm and} ~~ \frac{1-\beta}{2} B(\bp) =
 \begin{pmatrix} 0\\ 0\\ B_3 (\bp)\\ B_4 (\bp) \end{pmatrix} = B_\sg (\bp) v^\sg (0) ~, 
\lbl{3.30}
\ee
where $\sg=1,2$, and
\be
  u^1 (0) = \begin{pmatrix} 1\\ 0\\ 0\\ 0 \end{pmatrix}~~,
 ~~  u^2(0) = \begin{pmatrix} 0\\ 1\\ 0\\ 0 \end{pmatrix}~~,
     v^1 (0) = \begin{pmatrix} 0\\ 0\\ 1\\ 0  \end{pmatrix}~~,
     v^2 (0) = \begin{pmatrix}  0\\ 0\\ 0\\ 1 \end{pmatrix}~~,
\lbl{3.31}
\ee
and write
\be
   A_\sg (\bp) = \sqrt{\frac{m}{(2 \pi)^3 \om_\bp}} b_\sg (\bp) ~,~~~
    B_\sg (\bp) = i \sqrt{\frac{m}{(2 \pi)^3 \om_\bp}} d_\sg^* (\bp) .
\lbl{3.32}
\ee
Then we find that the wave function $\psi$ in Eq.\,(\ref{3.29}) is equal to the Dirac wave function
(\ref{3.17}), where $E = \om_\bp$.

\subsection{ An alternative choice of Hamiltonian}

Let us again consider the 4-component Klein-Gordon equation (\ref{3.2}) and its general solution
(\ref{3.5}) in which $\vphi \equiv \vphi^\alpha$, $\alpha =1,2,3,4$, are four real fields, and introduce
the 4-component field $\psi^\alpha$ according to Eq.\,(\ref{3.3}), but now with a different choice
of $h$, whose square is also $h^2 = m^2 + \p_i \p^i$. Instead of $h$ given in Eq.\,(\ref{3.1}), we will
take
\be
   h' = \beta \sqrt{m^2 + \p_i \p^i}~,~~~~  \beta = \begin{pmatrix} {\ul 1}&& 0\\ 0 && -{\ul 1} \end{pmatrix}~,
  ~~~  {\ul 1} = \begin{pmatrix} 1&& 0\\ 0 && 1 \end{pmatrix}  .
\lbl{3.33}
\ee
Inserting it into Eq.\,(\ref{3.3}), we obtain
\be
  \psi' = \sqrt{\frac{h'}{2}} \int \frac{\dd^3 \bp}{\sqrt{(2 \pi)^3 2 \om_\bp}} \left [(1+\beta) a(\bp)
  {\rm e}^{-p x} + (1-\beta) a^* (\bp) {\rm e}^{i p x} \right ] .
\lbl{3.34}
\ee
Using
\be
  \sqrt{h'} = \sqrt{\beta \om_\bx}~, \om_\bx = \sqrt{m^2 + \p_i \p^i}~,~~~
  \sqrt{\beta} = \begin{pmatrix} {\ul 1}&& 0\\ 0 && -{\ul i} \end{pmatrix} ,
\lbl{3.35}
\ee
we obtain
\be
\psi' = \int \frac{\dd^3 \bp}{\sqrt{(2 \pi)^3}} \left [\frac{1+\beta}{2} a(\bp)
{\rm e}^{-p x} + \frac{1-\beta}{2i} a^* (\bp) {\rm e}^{i p x} \right ] .
\lbl{3.36b}
\ee

Applying on the wave function $\psi'$ and on the Hamiltonian $h'$ the Foldy-Wouthuysen transformation\ci{FWtransformation}
\be
  {\rm e}^{i S} =  \frac{{\hat E} + m + \gam^i {\hat p}_i}{\sqrt{2 {\hat E}(m+{\hat E})}}~,
  ~~~{\rm e}^{-i S} =  \frac{{\hat E} + m - \gam^i {\hat p}_i}{\sqrt{2 {\hat E}(m+{\hat E})}},
\lbl{3.37}
\ee
where ${\hat E} = i \frac{\p}{\p t}$,  ${\hat p}_i = - i \p_i$, 
we obtain
$$
   \psi'' = {\rm e}^{-i S} \psi'  \hs{14cm}
$$
\be
  \hs{1mm} =\int \frac{\dd^3 \bp}{\sqrt{(2 \pi)^3}}\sqrt{\frac{1}{2 \om_\bp (m + \om_\bp)}}
    \left [ (\gam^\mu p_\mu + m) \frac{1+\beta}{2} a(\bp)
   {\rm e}^{-p x} + (-\gam^\mu p_\mu + m)\frac{1-\beta}{2i} a^* (\bp) {\rm e}^{i p x} \right ] .
   \lbl{3.38}
\ee
and
\be
   h'' = {\rm e}^{-i S} h' {\rm e}^{i S} = \beta (m - i \gam^i \p_i).
\lbl{3.38a}
\ee
The $\psi''$ in Eq.\,(\ref{3.38}) is equal to the Dirac wave function (\ref{3.17}) if we identify
\be
  \frac{1+\beta}{2} a(\bp) = \sqrt{\frac{m}{(2 \pi)^3 \om_\bp}}\, b_\sg (\bp) u^\sg (0)~,
  ~~~~\frac{1-\beta}{2} a^*(\bp) = i \sqrt{\frac{m}{(2 \pi)^3 \om_\bp}}\, d_\sg^* (\bp) v^\sg (0) .
\lbl{3.39}
\ee

The wave function $\psi''$ in Eq.\,(\ref{3.38}) is the probability amplitude of finding a particle
at position $\bx$ at a time $t$. The probability density is\footnote{
	We use $\left ( \frac{1+\beta}{2} \right )^2 = \frac{1+\beta}{2}$ and 
	$\left ( \frac{1-\beta}{2} \right )^2 = \frac{1-\beta}{2}$.}
\be
  \psi^\dg \psi = \int \frac{\dd^3 \bp \dd^3 \bp'}{(2 \pi)^3} 
  \left [ a^\dg (\bp) \frac{1+\beta}{2} a(\bp') {\rm e}^{i (\bp - \bp') \bx}
  + {a^*}^\dg (\bp) \frac{1-\beta}{2} a^*(\bp') {\rm e}^{-i (\bp - \bp') \bx} \right ].
\lbl{3.40}
\ee
By definition, it is positive. It containes a contribution of positive and negative energy states.
Integration over space gives
$$
  \int \dd^3 \bx \, \psi^\dg (x) \psi (x) = \int \dd^3 \bp \left [ a^\dg (\bp) \frac{(1+\beta)}{2}  \frac{(1+\beta)}{2} a(\bp)+
  a^{*\dg} (\bp) \frac{(1-\beta)}{-2i}  \frac{(1-\beta)}{2i} a^*(\bp) \right ]
$$
\be
  = \int \dd^3 \bp \, \left [ a_1^* (\bp) a_1 (\bp) + a_2^* (\bp) a_2 (\bp)
    + a_3^* (\bp) a_3 (\bp)  + a_4^* (\bp) a_4 (\bp) \right ] .
\lbl{3.41}
\ee

In view of the novel findings\ci{Pagani,Smilga1,Smilga2,Smilga3,Smilga4,PavsicFirence,PavsicStable,PavsicPUReview,Kapparulin,PavsicStumbling,Deffayet} concerning stability issues related to the presence of
negative energy states, it is quite legitimate to consider the Dirac wave function as
probability amplitude and its absolute square a probability density, without the usual caveat
that it makes sense only approximatively when negative energies can be neglected. 

\subsection{Probability density versus charge density}

The action for the spinor field (\ref{3.17}) is
\be
  I = \int \dd^4 x\, {\bar \psi} (i \gam^\mu \p_\mu - m ) \psi,
\lbl{3.24A}
\ee
where ${\bar \psi}= \psi^\dg \gam^0$. In general, the variation of an action that includes the variation
of the boundary is (see, e.g., Ref.\ci{BarutEM})
\be 
  \delta I = \int \dd^4 x\, \delta {\cal L} + \oint \dd \Sigma_\mu \, {\cal L} \delta x^\mu .
\lbl{3.25A}
\ee
For ${\cal L} = {\bar \psi} (i \gam^\mu \p_\mu - m ) \psi$ we have\footnote{For self-consistency we resume and
	adapt  here the standard procedure \`a la Noether.}
$$
  \delta I = \int \dd^4 x \, \p_\mu \left [\delta {\bar \psi} (i \gam^\mu \p_\mu - m) \psi +
  {\bar \psi} (i \gam^\mu \p_\mu - m) \delta \psi \right ] + \int \dd^4 x \, \p_\mu ({\cal L} \delta x) $$
\be
  =  \int \dd^4 x \, \p_\mu \left [ {\bar \psi} (i \gam^\mu \p_\mu - m) \psi 
  -(\p_\mu \bpsi \gam^\mu + \bpsi m) \delta \psi \right ]
  + \int \dd^4 x\, \p_\mu \left [ \bpsi i \gam^\mu \delta \psi + {\cal L} \delta x^\mu \right ].
\lbl{3.26A}
\ee
Here $\delta \psi = \psi'(x) - \psi(x)$ is the variation at a fixed point $x$. Introducing the total
variation ${\bar \delta} \psi = \psi'(x') - \psi(x)= \delta \psi + \p_\mu \delta x^\mu$ and assuming
that the equations of motion are satisfied, Eq.\,(\ref{3.26A}) becomes
\be
 \delta I = \int \dd^4 x \, \p_\mu \left [ {\bar \psi} i \gam^\mu {\bar \delta} \psi -
 (\bpsi i \gam^\mu \p_\nu \psi - {\cal L} {\delta^\mu}_\nu ) \delta x^\nu \right ] .
\lbl{3.27A}
\ee

Let us consider the case where $\delta x^\mu =0$ and ${\bar \delta} \psi =\delta \psi = i \alpha \psi$,
${\bar \delta}\bpsi = \delta \bpsi= - i \alpha \bpsi$, which is the infinitesimal form of the phase transformation
\be
  \psi' = {\rm e}^{i \alpha} \psi~,~~~~,  \bpsi' = {\rm e}^{-i \alpha} .
\lbl{3.28A}
\ee
Then
\be
  \delta I = \int \dd^4 x\, (-\alpha) \p_\mu (\bpsi \gam^\mu \psi ) .
\lbl{3.29A}
\ee
Because the action (\ref{3.24}) is invariant under the phase transformations (\ref{3.28A}),
the above expression vanishes and the current $ j^\mu = \bpsi \gam^\mu \psi$ is conserved.
The null component is $j^0 = \bpsi \gam^0 \psi = \psi^\dg \psi$; by definition it is positive.
If physicists were not puzzled by the presence of negative energies, $\psi^\dg \psi$ would have
been interpreted as probability density and $\bpsi \gam^i \psi$ as probability 3-current density.
But in Refs.\ci{Smilga1,Smilga2,Smilga3,PavsicFirence,PavsicStable,PavsicPUReview,Kaparulin,PavsicStumbling,Deffayet} it was found that negative energies are not problematic
at all, therefore we may interpret and we do interprete $j^\mu = \bpsi \gam^\mu \psi$ as
{\it the probability 4-current density}, whose null component is the probability densisty.

Where does then the electric charge come from? To obtain the electric charge, one has to double
the field $\psi$ and consider the action
\be
   I = \int \dd^4 x\, \left [{\bar \psi}_1 (i \gam^\mu \p_\mu - m ) \psi_1
    + {\bar \psi}_2 (i \gam^\mu \p_\mu - m ) \psi_2 \right ],
\lbl{3.42}
\ee
where $\psi_1 \equiv \psi_1^\alpha$,  $\psi_2 \equiv \psi_2^\alpha$, $\alpha=1,2,3,4$.
Each of those fields is composed \`a la (\ref{3.3}) in term of the {\it real} 4-component fields
\be
   \psi_1^\alpha = \sqrt{\frac{h}{2}}(\vphi_1^\alpha + i h^{-1} {\dot \vphi}_1^\alpha)~ ,~~~~~
   \psi_2^\alpha = \sqrt{\frac{h}{2}}(\vphi_2^\alpha + i h^{-1} {\dot \vphi}_2^\alpha) .
\lbl{3.43}
\ee
The expression (\ref{3.27A}) then generalizes to
\be
\delta I = \int \dd^4 x \, \p_\mu \left [ {\bar \psi}_1 i \gam^\mu {\bar \delta} \psi_1 +
{\bar \psi}_2 i \gam^\mu {\bar \delta} \psi_2 -
(\bpsi_1 i \gam^\mu \p_\nu \psi_1 + \bpsi_2 i \gam^\mu \p_\nu \psi_2- {\cal L} {\delta^\mu}_\nu ) \delta x^\nu \right ] .
\lbl{3.44}
\ee

There are two kinds of transformations that leave the action (\ref{3.42}) invariant:

1) A phase transformation of $\psi_n$:
\be
  \delta \psi_n = i \alpha \psi_n~,~~~n=1,2.
\lbl{3.45}
\ee
Then from (\ref{3.44}) we obtain two currents, $j_n^\mu = \bpsi_n \gam^\mu \psi_n$, one for
each component of the wave function $\psi_n$.

2) A rotation between $\psi_1$ and $\psi_2$:
\bear
  &&\psi'_1 = {\rm cos} \, \beta \,\psi_1 + {\rm sin} \,\beta \,\psi_2 , \nonumber \\
  &&\psi'_2 = - {\rm sin}\, \beta \,\psi_1 + {\rm cos} \,\beta \,\psi_2 , \nonumber \\
 && \Rightarrow \delta \psi_1 = \beta \psi_2~,~~~~\delta \psi_2 = - \beta \psi_1 .
\lbl{3.46}
\ear

Inserting the latter transformation into Eq.\,(\ref{3.44}), we obtain
\be
  \delta I = i \beta \int \dd^4 x\, \left [ \p_\mu (\bpsi_2 \gam^\mu \psi_1)
   -\p_\mu (\bpsi_1 \gam^\mu \psi_2) \right ],
\lbl{3.47}
\ee
which gives the conserved current
\be
  j_{12}^\mu = \bpsi_2 \gam^\mu \psi_1 -\bpsi_1 \gam^\mu \psi_2 .
\lbl{3.48}
\ee
Rewriting (\ref{3.42}) and (\ref{3.47}) in terms of the new variables (\ref{2.48}), we have
\be
  I = \int \dd^4 x \left [ {\bar \chi}_- ( i \gam^\mu \p_\mu - m) \chi_+ +
  	 {\bar \chi}_+( i \gam^\mu \p_\mu - m) \chi_- \right ],
\lbl{3.49a}
\ee
and
\be
\delta I = i \alpha \int \dd^4 x\,  \p_\mu ( {\bar \chi}_+ \gam^\mu \chi_+
  - {\bar \chi}_- \gam^\mu \chi_-) ,
\lbl{3.50}
\ee
where
\be
   {\bar \chi}_\pm = \chi_\pm^\dg \gam^0 ~,~~~~~
    \chi_\pm^\dg = \frac{\psi_1^\dg \pm {\bar i} \psi_2^\dg}{\sqrt{2}} .
\lbl{3.50a}
\ee
The conserved current is
\be
  {\tl j}^\mu =  {\bar \chi}_+ \gam^\mu \chi_+ - {\bar \chi}_- \gam^\mu \chi_- ,
\lbl{3.51}
\ee
where the components
\be
   {\tl j}^0 = \chi_+^\dg \chi_+ - \chi_-^\dg \chi_- ~,~~~~ {\rm and}  ~~~~~
   {\tl j}^i = \chi_+^\dg \gam^0 \gam^i \chi_+ - \chi_-^\dg \gam^0 \gam^i\chi_-
\lbl{3.52}
\ee
can be interpreted, respectively, as the { electric charge} and the {electric current density}.

Besides the imaginary unit $i$, which is the generator of the phase transformation
$\psi'_n = {\rm e}^{i \alpha} \psi_n$ (essentially a rotation between $\vphi_n$ and ${\dot \vphi}_n$),
we also have a distinct imaginary unit ${\bar i}$, which is the generator of rotation between
$\psi_1$ and $\psi_2$ that can be expressed as $\chi'_\pm = {\rm e}^{\mp {\bar i} {\tl \alpha}} \chi_\pm$.
We have thus two kinds of groups associated with the action (\ref{3.42}):
\begin{quote}
\  i) The group $U_i (1)$, acting on each of the fields $\psi_1$ and $\psi_2$.

  ii) The group $U_{\bar i}(1)$, acting on $\chi_\pm$.
\end{quote}
The first group, $U_i (1)$, is associated with the probability current density, whilst the second
group, $U_{\bar i}(1)$, is associated with (electric) charge current density.

Alternatively, to obtain electric charge, one can stay with a single field $\psi$ and consider
the extension of the action (\ref{3.24A}) to five dimensions\ci{Kaluza}). Then, from the
five-dimensional analog of Eq.\,(\ref{3.27A}), we have for the mixed components of the stress-energy
tensor:
\be
  {{\hat T}^\mu}_5 = i {\hat {\bar \psi}} {\hat \gam}^\mu \p_5 {\hat \psi}
  \propto e {\bar \psi}\gam^\mu \psi= j_e^\mu ,
\lbl{K1}
\ee
where ${\hat \psi} = {\rm e}^{-i p_5 x^5} \psi$ and where the electric charge is identified with $p_5$.
We see that the electric charge current density that comes form the 5th dimension is proportional
to the probability current density ${\bar \psi} \gam^\mu \psi$ of the usual Dirac theory in four
dimensions. The zero component $j^0 = e {\bar \psi} \gam^) \psi = e \psi^\dg \psi$ can now be positive or negative.

Traditionally, because of negative energies, the Dirac wave function was not considered as being physically viable\footnote{
Only in nonrelativistic approximation it was considered as making sense as a wave function whose
absolute square is the probability density. But in this study, we have shown that also the relativistic
Dirac wave function makes sense as a probability amplitude, though its decomposition in terms of
momenta contains positive and negative energies.}.
It has been maintained that one has to use quantum field theory to make sense of the Dirac
equation. As we will point out later, such a traditional approach makes sense anyway. It indeed leads
to the occurrence of charged particles (and antiparticles), without doubling the Dirac field $\psi$.
The theory considered here, in which we double the Dirac field, implies the existence of a
positively charged particle which is a partner of the electron, but not its antiparticle. 
 
\section{A formulation in terms of the phase space basis vectors: A step to quantum field theory}
\subsection{Vectors in symplectic and orthogonal spaces}

We started from the action (\ref{2.13}) for a real field $\vphi$ and wrote the phase space form
of the action (\ref{2.14}) which contains $\vphi$ and $\Pi = {\dot \vphi}$. In
Eqs.\,(\ref{2.15}),(\ref{2.16}) we introduced new variables $\psi$ and $\psi^*$ that satisfied the
Schr\"odinger equation (\ref{2.18a}) and its complex conjugate (\ref{2.18b}). The field $\vphi$ can be a
single component, two component or a 4-component {\it real} field $\vphi \equiv \vphi^\alpha (x)$,
and $\psi$ a complex field of the same number of components.

The field $\vphi^\alpha (\bx) \equiv \vphi^{\alpha (\bx)}$ and its canonically conjugated field
$\Pi^\alpha (\bx) \equiv \Pi^{\alpha (\bx)}$ are components of a vector $\Phi$ in phase space spanned
over an infinite dimensional basis $\lbrace k_{\alpha (\bx)}, {\bar k}_{\alpha (\bx)} \rbrace$.
We thus have
\be
  \Phi = \vphi^{\alpha (\bx)} k_{\alpha (\bx)} + \Pi^{\alpha (\bx)} {\bar k}_{\alpha (\bx)}.
\lbl{4.1}
\ee
Extending Eqs.\,(\ref{2.15}),(\ref{2.16}) to allow for 1, 2 or 4 components and for a generic
$h$, which can be $h = \om_\bx \equiv \sqrt{m^2 + {\hat p}^i {\hat p}_i}$, $h= \beta \om_\bx$ or
$h= \beta (m+ \gam^i {\hat p}_i)$, we have
\be
  \psi^{\alpha (\bx)} = \frac{1}{\sqrt{2}} \left (\sqrt{h} \vphi^{\alpha (\bx)}
   + \frac{i}{\sqrt{h}} \Pi^{\alpha (\bx)} \right )~,~~~~
   \psi^{*\alpha (\bx)} = \frac{1}{\sqrt{2}} \left (\sqrt{h} \vphi^{\alpha (\bx)}
   - \frac{i}{\sqrt{h}} \Pi^{\alpha (\bx)} \right ),
\lbl{4.2a}
\ee
the inverse transformation being
\be
  \vphi^{\alpha (\bx)} = \frac{1}{\sqrt{2}} \frac{1}{\sqrt{h}} (\psi^{\alpha (\bx)} + \psi^{*\alpha (\bx)}) ~,~~~~ \Pi^{\alpha (\bx)} = \frac{-i}{\sqrt{2}} \frac{1}{\sqrt{h}} (\psi^{\alpha (\bx)} - \psi^{*\alpha (\bx)}) .
\lbl{4.2}
\ee
Inserting the latter expressions into Eq.\,(\ref{4.1}), we obtain
\be
   \Phi = \psi^{\alpha (\bx)} q_{\alpha (\bx)}^\dg + \psi^{*\alpha (\bx)} q_{\alpha (\bx)} ,
\lbl{4.3}
\ee
where
\be
  q_{\alpha (\bx)} = \frac{1}{\sqrt{2}} \left ( \frac{1}{\sqrt{h}} k_{\al(\bx)} + i \sqrt{h} {\bar k}_{\alpha (\bx)} 
  \right )~,~~~~~~
    q_{\alpha (\bx)}^\dg = \frac{1}{\sqrt{2}} \left ( \frac{1}{\sqrt{h}} k_{\al(\bx)} - i \sqrt{h} {\bar k}_{\alpha (\bx)} 
   \right ), \lbl{4.4}
\ee

A vector in phase space can thus be expressed either in terms of the basis vectors $k_{\alpha(\bx)}$,
${\bar k}_{\alpha(\bx)}$ acording to (\ref{4.1}), or it can be expressed in terms of the basis
vectors $q_{\alpha (\bx)}$, $q_{\alpha (\bx)}^\dg$ according to (\ref{4.3}). In a compact notation
we have
\be
  \Phi = \vphi^{i \alpha (\bx)} k_{i \alpha (\bx)} = \psi^{i \alpha (\bx)} q_{i \alpha (\bx)}~ ,
  ~~~ i = 1,2,
\lbl{4.8}
\ee
where
\be
  \vphi^{i \alpha (\bx)} = (\vphi^{\alpha (\bx)},\Pi^{\alpha (\bx)}) ~,
  ~~~ \psi^{i \alpha (\bx)} = (\psi^{\alpha (\bx)}, \psi^{*\alpha (\bx)}) ,
\lbl{4.8a}
\ee
\be
  k_{i \alpha (\bx)} = (k_{\alpha (\bx)}, {\bar k}_{\alpha (\bx)})~, 
  ~~~~   q_{i \alpha (\bx)} = (q_{\alpha (\bx)}^\dg, q_{\alpha (\bx)}) .
\lbl{4.8b}
\ee

For the basis vectors $k_{i \alpha (bx)}$ we take the generators of the Clifford algebra\footnote{
Here we adopt the methods of geometric algebra as introduced by Hestenes\ci{Hestenes1,Hestenes2},
and extend it to infinite-dimensional phase space (see \ci{PavsicOrthoSymp}).}
of the symplectic or orthogonal phase space.

We distinguish two cases:

a) {\it Symplectic case}, in which the inner product is given by
\be
  k_{i \alpha (\bx)} \wg k_{j \beta (\bx')} = \frac{1}{2} [ k_{i \alpha (\bx)}, k_{j \beta (\bx')}]
  = \rho_{i \alpha (\bx) j \beta (\bx')} ,
\lbl{4.6a}
\ee
where
\be
   \rho_{i \alpha (\bx) j \beta (\bx')} =  J_{i \alpha (\bx) j \beta (\bx')}
   = \begin{pmatrix}
   	   0 &  \delta_{\alpha (\bx) \beta (\bx')} \\
   	   - \delta_{\alpha (\bx) \beta (\bx')} & 0
   	  \end{pmatrix}
     =  \epsilon_{ij} \otimes \delta_{\alpha (\bx) \beta (\bx')}
\lbl{4.7a}
\ee 
is the symplectic metric. Here $\epsilon_{ij} = - \epsilon_{ji}$ and $\delta_{\alpha (\bx) \beta (\bx')}
= \delta_{\alpha \beta} \delta^3 (\bx-\bx')$.
The basis vectors $q_{\alpha(\bx)}$, $q_{\alpha(\bx)}^\dg$ then satisfy
\be
  q_{\alpha(\bx)} \wg q_{\beta(\bx')}^\dg = \frac{1}{2} [q_{\alpha(\bx)}, q_{\beta(\bx')}^\dg]
  = \delta_{\alpha (\bx) \beta (\bx')} .
\lbl{4.7b}
\ee
\be
   [q_{\alpha(\bx)}, q_{\beta(\bx')}] = 0~,
~~~~ [q_{\alpha(\bx)}^\dg, q_{\beta(\bx')}^\dg ] = 0 ,
\lbl{4.7c}
\ee 
compactly,
\be
  q_{i \alpha (\bx)} \wg q_{j \beta (\bx')} = \rho'_{i \alpha (\bx) j \beta (\bx')} =
  \begin{pmatrix}
  	0 & - \delta_{\alpha (\bx) \beta (\bx')} \\
  	 \delta_{\alpha (\bx) \beta (\bx')} & 0
  \end{pmatrix}	.
\lbl{4.7eee}
\ee

b) {\it Orthogonal case}, in which the inner product is 
\be
    k_{i \alpha (\bx)} \cdot k_{j \beta (\bx')}
     = \frac{1}{2} \lbrace k_{i \alpha (\bx)}, k_{j \beta (\bx')} \rbrace
   = \rho_{i \alpha (\bx) j \beta (\bx')} ,
\lbl{4.6b}
\ee
where 
\be
  \rho_{i \alpha (\bx) j \beta (\bx')}
  = \begin{pmatrix}
  	h \delta_{\alpha (\bx) \beta (\bx')} & 0 \\
  	0 & \frac{1}{h} \delta_{\alpha (\bx) \beta (\bx')}
  \end{pmatrix}
\ee
is the orthogonal metric. The basis vectors $q_{\alpha(\bx)}$, $q_{\alpha(\bx)}^\dg$ now
satisfy
\be
q_{\alpha(\bx)} \cdot q_{\beta(\bx')}^\dg = \frac{1}{2} \lbrace q_{\alpha(\bx)}, q_{\beta(\bx')}^\dg\rbrace
= \delta_{\alpha (\bx) \beta (\bx')} .
\lbl{4.7ee}
\ee
\be
   \lbrace q_{\alpha(\bx)}, q_{\beta(\bx')} \rbrace = 0~,
   ~~~~\lbrace q_{\alpha(\bx)}^\dg, q_{\beta(\bx')}^\dg \rbrace = 0 ,
\lbl{4.7d}
\ee
which in the compact notation reads:
\be
q_{i \alpha (\bx)} \cdot q_{j \beta (\bx')} = q'_{i \alpha (\bx) j \beta (\bx')} =
\begin{pmatrix}
	0 &  \delta_{\alpha (\bx) \beta (\bx')} \\
	\delta_{\alpha (\bx) \beta (\bx')} & 0
\end{pmatrix}	.
\lbl{4.7e}
\ee
  
Explicitly, the symplectic relation (\ref{4.6a}) gives
\be
  \frac{1}{2} [k_{\alpha (\bx)},{\bar k}_{\beta (\bx')} ] = i \delta_{\alpha (\bx) \beta (\bx')}~,
  ~~~~  [k_{\alpha (\bx)},{k}_{\beta (\bx')} ] = 0~,
  ~~~~ [{\bar k}_{\alpha (\bx)},{\bar k}_{\beta (\bx')} ] = 0 .
\lbl{4.9}
\ee
If we identify
\be
   \frac{k_{\alpha (\bx)}}{\sqrt{2}} = {\hat \vphi}_\alpha (\bx) ~,~~~~~
   \frac{{\bar k}_{\alpha (\bx)}}{\sqrt{2}} = i{\hat \Pi}_\alpha (\bx) ,
\lbl{4.10}
\ee
we see that Eqs.\,(\ref{4.9}) become the equal time commutation relations for the bosonic
quantum field ${\hat \vphi}_\alpha$ and its conjugated momentum ${\hat \Pi}_\alpha$:
\be
  [{\hat \vphi}_\alpha (\bx), {\hat \Pi}_\beta (\bx')] = i \delta_{\alpha \beta} \delta^3 (\bx-\bx')~,
~~~~~[{\hat \vphi}_\alpha (\bx), {\hat \vphi}_\beta (\bx')] = 0~,
~~~~[{\hat \Pi}_\alpha (\bx), {\hat \Pi}_\beta (\bx')] = 0 .
\lbl{4.11}
\ee
The symplectic basis vectors $k_{\alpha (\bx)}$, ${\bar k}_{\alpha (\bx)}$, satisfying (\ref{4.9}),
are thus---up to the factor $1/\sqrt{2}$---just the bosonic quantum fields.

Analogously, in the {\it orthogonal case}, Eqs.\,(\ref{4.7c}),\,(\ref{4.7d}) are the
anticommutation relations for the fermionic field operators ${\hat \psi}_{\alpha(\bx)}
 = \frac{q_{\al(\bx)}}{\sqrt{2}}$ and ${\hat \psi}_{\alpha(\bx)}^\dg
  = \frac{q_{\al(\bx)}^\dg}{\sqrt{2}}$.

A phase space vector can be expresses either in the form (\ref{4.8}) or in the form
\be
   \Phi = \vphi_{i \al(\bx)} k^{i \al (\bx)} = \psi_{i \al(\bx)} q^{i \al (\bx)},
\lbl{4.15}
\ee
where
\be
   k^{i \al (\bx)} = \rho^{i \al (\bx) j \beta (\bx')}  k_{j \beta (\bx')} ~~~~{\rm and}
   ~~~~ q^{i \al (\bx)} = \rho'^{i \al (\bx) j \beta (\bx')}  q_{j \beta (\bx')} .
\lbl{4.16}
\ee
By using the appropriate inverse metrix we find that both in the symplectic and orthogonal case
we have
\be
  q^{i \al (\bx)} = \left ( \frac{1}{\sqrt{2}} \left ( \sqrt{h} k^{1 \al (\bx)} +
  \frac{i}{\sqrt{h}} k^{2 \al (\bx)} \right ), \frac{1}{\sqrt{2}} 
  \left (  \sqrt{h} k^{1 \al (\bx)} -\frac{i}{\sqrt{h}} k^{2 \al (\bx)} \right ) \right ),
\lbl{4.17}
\ee
\be
\psi_{i \al (\bx)} = \left ( \frac{1}{\sqrt{2}} \left ( \frac{1}{\sqrt{h}} \vphi_{1 \al (\bx)} -
i{\sqrt{h}} \vphi_{2 \al (\bx)} \right ), \frac{1}{\sqrt{2}} 
\left ( \frac{1}{\sqrt{h}} \vphi_{1 \al (\bx)} +
i{\sqrt{h}} \vphi_{2 \al (\bx)} \right )  \right ).
\lbl{4.18}
\ee

\subsection{Constructing quantum field theory out of phase space basis vectors}

A phase space basis vector (\ref{4.3}) describes a single-particle state, the wave function and its
complex conjugate being $\psi^{\al (\bx)}$ and $\psi^{*\al (\bx)}$, defined in Eq.\,(\ref{4.2a}).
The probability density is given by $\psi^* \psi$. However, if we take the inner product of the vector
$\Phi$ with itself, then we obtain
\be
  \langle \Phi \Phi \rangle_S = \Phi \wg \Phi =0~~~~~\text {for symplectic vector},
\lbl{4.19}
\ee
\be
  \langle \Phi \Phi \rangle_S = \Phi \cdot \Phi = {\hat h} \vphi^2 + {\hat h}^{-1} \Pi^2 =
  2 \psi^* \psi~~~~\text{for orthogonal vector}.
\lbl{4.20}
\ee

Taking into account that a vector $\Phi$ is the sum of two vectors
\be
  \Phi = {\tl \Psi} + {\tl \Psi}^\dg ~,~~~~~{\tl \Psi} = \psi^{\al (\bx)} q_{\al (\bx)}^\dg~, 
  ~~~~~ {\tl \psi}^\dg = \psi^{*\al (\bx)} q_{\al (\bx)},
\lbl{4.21}
\ee
we have
\be
  \langle {\tl \Psi}^\dg  {\tl \Psi} \rangle_S 
  = \psi^{\al (\bx)} \psi^{\beta (\bx')} \langle q_{\al (\bx)} q_{\beta (\bx')} \rangle_S
 = \psi^{*\al (\bx)} \psi^{\beta (\bx')} \delta_{\al (\bx) \beta (\bx')},
\lbl{4.22}
\ee
where
\be
   \langle q_{\al (\bx)} q_{\beta (\bx')} \rangle_S  =  
   \begin{cases}
    q_{\al (\bx)} \wg q_{\beta (\bx')} = \delta_{\al (\bx) \beta (\bx')} ~~
   \text{for symplectic case},\\
   q_{\al (\bx)} \cdot q_{\beta (\bx')} = \delta_{\al (\bx) \beta (\bx')}   ~~\text{for orthogonal case}.
   \end{cases}
\lbl{4.23}
\ee

Introducing a vacuum state $\Omega$ such that
\be
  q \Omega = 0~,~~~~\Omega^\dg q^\dg =0 ,
\lbl{4.24}
\ee
let us define a vector $\Psi$ according to
\be
  \Psi = \Phi \Omega = \psi^{\al (\bx)} q_{\al (\bx)} \Omega.
\lbl{4.25}
\ee
Then we have
$$
  \langle \Psi^\dg \Psi \rangle_S = \langle \Om^\dg \Phi \Phi \Om \rangle_S =
  \langle \Omega^\dg \psi^{*\al (\bx)} q_{\al (\bx)} q_{\beta (\bx')}^\dg \psi^{\beta (\bx')} \Omega \rangle_S$$
\be
  \hs{2cm}= 2 \langle \Omega^\dg \Omega \rangle_S \psi^{*\al (\bx)} \psi^{\beta (\bx')} \delta_{\al (\bx) \beta (\bx')} .
 \lbl{4.26}
\ee

Expressed in terms of the operators $a_{\al(\bx)} = q_{\al(\bx}/\sqrt{2}$, $a_{\al(\bx)}^\dg 
= q_{\al(\bx}^\dg \sqrt{2}$ and the vacuum $\vac = \sqrt{2} \Omega$,
$\langle 0| = \sqrt{2} \Omega^\dg$, $\langle 0| \vac = 2 \Omega^\dg \Omega =1$,
the inner products then reads
$$
  \langle \Psi^\dg \Psi \rangle_S = \vacc \psi^{*\al(\bx} q_{\al(\bx)} q^\dg_{\beta(\bx')} \psi^{\beta (\bx)} \vac
  = \vacc \psi^{*\al(\bx)} (\delta_{\al(\bx) \beta(\bx')} \pm q_{\beta(\bx')} q_{\al(\bx)}^\dg ) \psi^{\beta(\bx')} \vac
$$
\be
    \hs{1cm} =\psi^{*\al(\bx)} \psi^{\beta(\bx')} \delta_{\al(\bx) \beta(\bx')}
   \equiv \int \dd^3 \bx\, \dd^3 \bx' \, \psi^{* \al} (\bx) \psi^{\beta} (\bx') \delta_{\al \beta} \delta^3 (\bx-\bx') .
\lbl{4.26a}
\ee
Here the plus sign refers to the symplectic and the minus sign to the orthogonal case.

Let us now consider the inner product of the vector $\Phi$ with its time derivative ${\dot \Phi}$.
Using Eqs.\,(\ref{4.1}),(\ref{4.3},(\ref{4.7a}) and (\ref{4.7b}), we obtain
$$
  \langle \Phi {\dot \Phi} \rangle_S = \Phi \wg {\dot \Phi} 
  	= - i \left ( \Pi^{\alpha('bx)} {\dot \vphi}^{\beta(\bx)} 
  		-  \vphi^{\al(\bx)} {\dot \Pi}^{\beta (\bx')} \right ) \delta_{\al(\bx) \beta(\bx')}$$
\be
  	= \left ( \psi^{\al(\bx)} {\dot \psi}^{*\beta(\bx)} - \psi^{*\al(\bx)} {\dot \psi}^{\beta(\bx')}   \right )
  	 \delta_{\al(\bx) \beta(\bx')}
\lbl{4.27}
\ee
for {\it the symplectic case}, and
$$
\langle \Phi {\dot \Phi} \rangle_S = \Phi \cdot {\dot \Phi} 
  = \vphi^{\al(\bx)} h_{\al(\bx) \beta(\bx')} {\dot \vphi}^{\beta(\bx')} 
  + \Pi^{\al(\bx)} h_{\al(\bx) \beta (\bx')}^{-1} {\dot \Pi}^{\beta(\bx')}$$
$$
 \hs{3.5cm} = \frac{\dd}{\dd t} \left ( \vphi^{\al(\bx)} h_{\al(\bx') \beta(\bx)} \vphi^{\beta(\bx')} 
  + \Pi^{\al(\bx)} h_{\al(\bx) \beta(\bx')}^{-1} \Pi^{\beta (\bx')} \right )$$
$$
  =\psi^{\al(\bx)} \delta_{\al(\bx) \beta(\bx')} {\dot \psi}^{*\beta(\bx')} +
  {\dot \psi}^{\al(\bx)} \delta_{\al(\bx) \beta(\bx')} \psi^{*\beta(\bx')}$$
\be
 \hs{3cm} = \frac{\dd}{\dd t} \left ( \psi^{\al(\bx)} \psi^{*\beta (\bx')} \delta_{\al(\bx) \beta (\bx')} \right ),
\lbl{4.28}
\ee
for {\it the orthogonal case}.
  	
If instead of the vector $\Phi$ we take the $\Psi$ of Eq.\,(\ref{4.25}), then we can define the
following inner product
$$
  \langle \Psi^\dg {\dot \Psi} \rangle = \langle 0|\psi^{*\al(\bx)} a_{\al(\bx)} a^\dg_{\beta(\bx')}
   {\dot \psi}^{\beta (\bx')} \vac
    = \psi^{*\al(\bx)} {\dot \psi}^{\beta(\bx')} \delta_{\al(\bx) \beta(\bx')} $$
\be
  = \frac{i}{2} (\Pi {\dot \vphi} - \vphi {\dot \Pi})
   + \frac{1}{4} \frac{\dd}{\dd t} (\vphi h \vphi + \Pi h^{-1} \Pi ),
\lbl{4.29}
\ee
  
The action can then be rewritten as
\be
 I = \int \dd t \left ( i \langle \Psi^\dg {\dot \Psi} \rangle_S - H \right )
  = \int \dd t \, \left ( i \psi^* {\dot \psi} - H \right ) 
  = \int \dd t \left [\frac{1}{2} (\Pi {\dot \vphi} - \vphi {\dot \Pi}) - H \right ].	
\lbl{4.30}
\ee

The Hamiltonian is a generalization of the expression (\ref{2.21}) and (\ref{2.22}):
\be
  H= \frac{1}{2} \psi^{i \al(\bx)} h_{i \al(\bx)j \beta(\bx')} \psi^{j \beta(\bx')}
  \equiv \frac{1}{2} \psi^i h_{ij} \psi^j ,
\lbl{4.31}
\ee
where $h_{i \al(\bx)j \beta(\bx')} \equiv h_{ij}$ is an appropriate matrix to be determined.

To simplify the notation we have omotter the double index $\al(\bx)$. In such a simplified
notation we have
\be
  \Phi = \vphi^i k_i = \vphi'^i k'_i \equiv \psi^i q_i = (\psi q^\dg + \psi^* q ) ,
\lbl{4.32}
\ee
\be
  \Psi = \Phi \Omega =  (\psi q^\dg + \psi^* q ) \Omega = \psi a^\dg \vac,
\lbl{4.33}
\ee
where
\be
  \Om = \frac{1}{\sqrt{2}}\vac~,  ~~~~~~~ \frac{q}{\sqrt{2}} = a~,
  	~~~~~~\frac{q^\dg}{\sqrt{2}} = a^\dg ,
\lbl{4.34}
\ee
\be
   k_i = (k, {\bar k})~,~~~~~~k'_i \equiv q_i .
\lbl{4.35}
\ee

The inner porduct is
\begin{equation}
	\lbrace k_i k_j \rbrace_S =
	\begin{cases}
		k_i \wg k_j = \rho_{ij} = \begin{pmatrix}
			0 & 1\\
			-1 & 0
		\end{pmatrix}
			 & \text{symplectic case}\\
	  	k_i \cdot k_j = \rho_{ij} = \begin{pmatrix}
	  	h & 0\\
	  	0 & h^{-1}
	  \end{pmatrix}
		 & \text{orthogonal case}
	\end{cases}
\lbl{4.36}       
\end{equation}   
 \begin{equation}
 	\lbrace q_i q_j \rbrace_S =
 	\begin{cases}
 		q_i \wg q_j = \rho'_{ij} = \begin{pmatrix}
 			0 & 1\\
 			-1 & 0
 		\end{pmatrix}
 		& \text{symplectic case}\\
 		q_i \cdot q_j = \rho'_{ij} ~~= \begin{pmatrix}
 			0 & 1\\
 			1 & 0
 		\end{pmatrix}
 		& \text{orthogonal case}
 	\end{cases}
 	\lbl{4.37}       
 \end{equation}   
Here $k_i \equiv k_{i \al(\bx)} = (k_{\al (\bx)},{\bar k}_{\al (\bx)}$, $q_i \equiv q_{i \al (\bx}$,
$\vphi_i \equiv \vphi_{i \al(\bx)}$, $\psi_i \equiv \psi_{i \al (\bx)}$, $\rho_{ij} \equiv
\rho_{i \al (\bx) j \beta(\bx')}$. Indices are raised by the inverse matrix $\rho^{ij} \equiv
\rho^{i \al (\bx) j \beta(\bx')}$, satisfying
\be
  \rho^{ik} \rho_{kj} = \rho'^{ik} \rho'_{kj} = {\delta^i}_j = \langle k^i k_j \rangle_S
  = \langle q^i q_j \rangle_S .
\lbl{4.38}
\ee
In the orthogonal case we have 
\be
  q^i = (q,q^\dg)~,~~~~q_i = (q^\dg,q)~,~~~~
\rho'_{ij} 
 = \begin{pmatrix}
  0 & 1\\
  1 & 0
\end{pmatrix} ~,~~~
  \rho'^{ij}
   = \begin{pmatrix}
  	0 & 1\\
  	1 & 0
  \end{pmatrix} ,
\lbl{4.38a}
\ee
\be
  \psi^i = (\psi,\psi^*)~,~~~~~\psi_i = (\psi^*,\psi) .
\lbl{4.38b}
\ee

The Hamiltonian (\ref{4.31}) can be written as
\be
  H= \frac{1}{2} \psi^i {\delta_i}^k h_{k\ell} {\delta^\ell}_j \psi^j
  = \frac{1}{8} \psi^i (q_i q^k + q^k q_i) h_{k \ell} (q^\ell q_j + q_j q^\ell) \psi^j .
\lbl{4.39}
\ee

Because the Hamiltonian is a scalar, we have $H = H \langle 0| \vac = \langle 0|H \vac$. Therefore,
Eq.\,(\ref{4.39}) can be sandwiched between the vacuum state:
\be
   H= 
  = \frac{1}{8} \langle 0 | \psi^i (q_i q^k + q^k q_i) h_{k \ell} (q^\ell q_j + q_j q^\ell) \psi^j \vac.
\lbl{4.40}
\ee
It turns out that the above expression is equal to
\be
  H = \frac{1}{8} \langle 0| \psi^i q_i q^k h_{k\ell} q^\ell q_j \psi^j \vac ,
\lbl{D1}
\ee
which in view of eqs.(\ref{4.38a}),(\ref{4.38b}) and (\ref{4.24}) is equal to
\be
  H=\frac{1}{8} \langle 0| \psi^* q q^k h_{k\ell} q^\ell q^\dg \psi \vac =
  \frac{1}{2} \langle 0| \psi^* a a^k h_{k \ell} a^\ell a^\dg \psi \vac .
\lbl{D2}
\ee
  
Expressed in terms of $a_i$, $a^i$, the Hamiltonian (\ref{D1}) thus reads
\be
  H = \frac{1}{2} \langle 0|\psi^* a\, a^i h_{ij} a^j a^\dg \psi \vac
    = \langle \psi|{\hat H}|\psi \rangle ,
\lbl{4.45}
\ee
where $a^i = (a,a^\dg)$, and
\be
  |\psi \rangle = \psi a^\dg \vac \equiv \int \dd^3 \bx \, \psi (\bx) a^\dg (\bx) \vac~ ,~~~~
 \langle \psi| = \langle 0| \psi^* a \equiv \langle 0| \int \dd^3 \bx\, \psi^* (\bx) a(\bx) ,
\lbl{4.46}
\ee
and
\be
 {\hat H} = \frac{1}{2} a^i h_{ij} a^j \equiv \frac{1}{2} a^{i \al(\bx)} h_{i \al(\bx) j \beta(\bx')} a^{j (\bx)}
\lbl{4.47}
\ee
is the Hamilton operator.
In the last step in the above equation we included also the spinor index $\al,\beta= =1,2,3,4$ and
the spatial coordinates $\bx$ (written as a continuous index).

Explicitly, the Hamilton operator reads
\be
{\hat H} = \frac{1}{2} \int \dd \bx'\, \dd \bx'' \left ( a^\alpha (\bx')
(h_{12})_{\alpha \beta} (\bx,\bx'') a^{\dg \beta} (\bx'') +
a^{\dg \beta} (h_{21})_{\beta \alpha} (\bx'',\bx') a^\alpha (\bx') \right ) .
\lbl{D3}
\ee

We have derived the expression (\ref{4.45})--(\ref{4.47}) for the orthogonal case in which $a$, $a^\dg$ are
fermionic creation/annihilation operators. The same expression comes out also in the symplectic case,
in which $a$, $a^\dg$ are bosonic operators and $h_{ij}$ an appropriate matrix, e.g.,
$h_{ij} \equiv h_{i(\bx)j(\bx')} = \om_\bx \delta^3 (\bx-\bx')$.
	
Let us now identify
\be
  a^{i \al(\bx)} \equiv {\hat \psi}^{i \al (\bx)} = ({\hat \psi}^{\al(\bx)}, {\hat \psi}^{\dg \al(\bx)}),
\lbl{4.48}
\ee
where ${\hat \psi}^{i \al (\bx)}$ is the Dirac spinor field operator, and take the usual expression
for the Hamilton operator,
\be
  {\hat H} = \int \dd^3 \bx\, {\hat \psi}^\al (\bx) (\beta m - i \p_i)_{\al \beta} {\hat \psi}^\beta (\bx) .
\lbl{4.49}
\ee
Using
\be
  i \p_0 {\hat \psi} = (\beta m - i \al^i \p_i ) \hpsi,
\lbl{4.49a}
\ee
we have
\be
  {\hat H} = \int \dd^3 \bx \, \hpsi^\dg i \p_0 \hpsi .
\lbl{4.50}
\ee
Inserting the conventional expression
\be
  {\hat \psi} = \sum_\sg \frac{\dd^3 \bp}{(2 \pi)^3} \frac{m}{E}
  \left ( b_\sg (\bp) u_\sg (\bp) {\rm e}^{-i p x} + d_\sg^\dg (\bp) v_\sg (\bp) {\rm e}^{i p x} \right ),
\lbl{4.51}
\ee
which gives $\lbrace {\hat \psi}^\alpha (\bx), {\hat \psi}^\beta (\bx') \rbrace =
\delta^{\alpha \beta} \delta^3 \delta (\bx - \bx')$,  if
\be
  \lbrace b_\sg^{\dg \alpha} (\bp), b_{\sg'}^{\dg \beta} (\bp') \rbrace
  = (2 \pi)^3 \frac{E}{m} \delta^{\alpha \beta} \delta^3 (\bp - \bp')
  = \lbrace d_\sg^{\dg \alpha} (\bp), d_{\sg'}^{\dg \beta} (\bp') \rbrace ,
\lbl{4.50a}
\ee
and using $u_\sg^\dg (\bp) u_{\sg'} (\bp') = v_\sg^\dg (\bp) v_{\sg'} (\bp')
 = \frac{E}{m} \delta_{\sg \sg'}$,
we obtain the usual expression for the Hamilton operator,
\be
  {\hat H} =  \sum_\sg \frac{\dd^3 \bp}{(2 \pi)^3} m \left (b_\sg^\dg (\bp) b_\sg (\bp)
  - d_\sg (\bp) d_\sg^\dg (\bp) \right ).
\lbl{4.52}
\ee
The same expression we also obtain if instead of (\ref{4.50}) we use the expression
\be
  {\hat H} = \frac{1}{2} \int \dd^3 \bx \, \left ( \hpsi^\dg i \p_0 \hpsi
  - i \p_0 {\hat \psi} {\hat \psi}^\dg \right ) ,
\lbl{4.53}
\ee
which gives
\be
  \frac{1}{2}  \sum_\sg \frac{\dd^3 \bp}{(2 \pi)^3} m \left ( b_\sg^\dg (\bp) b_\sg (\bp)
  - d_\sg (\bp) d_\sg^\dg (\bp)  - b_\sg (\bp) b_\sg^\dg (\bp) + d_\sg^\dg (\bp) d_\sg (\bp) \right ) .
\lbl{4.54}
\ee
Because of the anticommutation relations (\ref{4.50a}), the zero point energies cancel out, and
we obtain Eq.\,(\ref{4.52}).
 
Using (\ref{4.49a}), the Hamilton operator (\ref{4.53}) becomes
\be
  {\hat H} = \frac{1}{2} \int \dd^3 \bx\, \left [{\hat \psi}^\dg (\beta m - i \alpha^i \p_i) {\hat \psi}
  -(\beta m - i \alpha^i \p_i) {\hat \psi} {\hat \psi}^\dg \right ] .
\lbl{4.55}
\ee
This can be rewritten into the form
\be
  {\hat H} =\frac{1}{2} {\hat \psi}^{i \alpha(\bx)} h_{i \alpha(\bx)j \beta (\bx')}
  	{\hat \psi}^{j \beta(\bx')} ,
\lbl{4.56}
\ee
where
\be
   h_{i \alpha(\bx)j \beta (\bx')} =
   \begin{pmatrix}
   	0 & - (m \beta^* - i\alpha^{*i} \p_i)_{\alpha \beta} \delta^3 (\bx-\bx')\\
   	( m \beta- i\alpha^{i} \p_i)_{\alpha \beta} \delta^3 (\bx-\bx')&    0
   	\end{pmatrix}
\lbl{4.56a}
\ee
Here we have used the Hermiticity property
$h_{i \alpha(\bx)j \beta(\bx')}^\dg = h_{i \alpha(\bx)j \beta(\bx')} = h_{j \beta(\bx')i \alpha(\bx)}^*$.

The equality of the expressions (\ref{4.49}) and (\ref{4.55}) implies in the short hand notation that
\be
  {\hat H} = a^\dg h_{21} a = \frac{1}{2} (a h_{12} a^\dg + a^\dg h_{21}a)
  = \frac{1}{2} a^i h_{ij} a^j ,
\lbl{4.57}
\ee
and
$$
  \langle \Psi|{\hat H} |\Psi \rangle = \langle 0| \psi^* a a^\dg h_{21} a a^\dg \psi \vac =
  \psi^* h_{21} \psi  \hs{6cm} $$
\be
   \hs{1cm} = \frac{1}{2} \langle 0| \psi^* a a^i h_{ij} a^j a^\dg \psi \vac
    = \frac{1}{2} (\psi^* h_{21} \psi + \psi h_{12} \psi^* ) = \frac{1}{2} \psi^i h_{ij} \psi^j .
\lbl{4.58}
\ee
This shows that the classical Hamiltonian (\ref{4.31}) is equal to the expectation value (\ref{D2})
of the quantum field operator ${\hat H}$ in a single-particle state (\ref{4.46}), and confirms the equality
of the expressions (\ref{4.40}) and (\ref{D1}).

The evolution of a quantum state, $|\psi \rangle = \psi^{\alpha (\bx)} a^\dg_{\alpha (\bx)} \vac$,
is determined by the Hamilton operator as
\be
  i \frac{\dd |\Psi \rangle}{\dd t} = {\hat H} |\Psi \rangle  .
\lbl{4.59}
\ee

Let us first assume that {\it the time dependence is in the wace function}. Then we have
\be
i \frac{\dd |\Psi \rangle}{\dd t} = i {\dot \psi}^{\alpha(\bx)} a_{\alpha(\bx)}^\dg \vac = {\hat H} |\Psi \rangle .
\lbl{4.60}
\ee
For the Hamilton operator let us take Eq.\,(\ref{4.47}) which for the free Dirac field can be written as
\be
  {\hat H} = a^{\dg \alpha(\bx)} h_{\alpha (\bx) \beta(\bx')} a^{\beta(\bx')} .
\lbl{4.60a}
\ee
Multiplying Eq.\,(\ref{4.60}) from the left by $\langle 0|a^{\beta(\bx')}$, we find
\be
  i {\dot \psi}^{\alpha (\bx)} = {h^{\alpha (\bx)}}_{\beta(\bx')} \psi^{\beta(\bx')}
  = \lbrace \psi^{\alpha (\bx)},H \rbrace_{P.B.} ,
\lbl{4.61}
\ee
where\footnote{In the case of a scalar field, it is ${h^{\alpha (\bx)}}_{\beta(\bx')} =
\sqrt{m^2 - \p^i \p_i} {\delta^\alpha}_\beta \delta^3 (\bx-\bx')$}
\be
 { h^{\alpha (\bx)}}_{\beta(\bx')} = {(\beta m - i \alpha^i \p_i )^\alpha}_\beta  \, \delta^3 (\bx-\bx')
\lbl{4.62}
\ee
and
\be
  H= \psi^{*\alpha(\bx)}  h_{\alpha (\bx) \beta(\bx')} \psi^{\beta(\bx')}
\lbl{4.63}
\ee
is the classical Hamiltonian (i.e., the Hamiltonian of the ``first quantized'' field).

Alternatively, if we assume the {\it the time dependence is in the operators} $a(\bx)$, $a^{\dg} (\bx)$,
then we have
\be
i \frac{\dd |\Psi \rangle}{\dd t} = i {\psi}^{\alpha (\bx)} {\dot a}_{\alpha(\bx)}^\dg \vac = {\hat H} |\Psi \rangle,
\lbl{4.64}
\ee
which gives
\be
  i \psi^{\alpha(\bx)} {\dot a}_{\alpha(\bx)}^\dg \vac =
  a^{\dg \beta(\bx')} h_{\beta(\bx') \alpha(\bx)} \psi^{\alpha(\bx)} \vac .
\lbl{4.65}
\ee
Because the latter equation holds for any $\psi^{\alpha(\bx)}$, it follows that
\be
  i {\dot a}_{\alpha(\bx)}^\dg = a^{\dg \beta (\bx')} h_{\beta(\bx')\alpha(\bx)}
  = [{\hat H},a_{\alpha(\bx)}^\dg] ,
\lbl{4.66}
\ee
which is the Heisenberg equation of motion.

The second equality in the latter equation can be straightforwardly derived by using
$[BC,A]=B[C,A] + [B,A]C$ for the bosonic operators, and $[BC,A]=B[C,A] - [B,A]C$
for the fermionic operators. In the case of the Dirac field, we can also proceed as
follows. Namely, using the explicit notation for the Hamilton operator, 
\be
  {\hat H} = \int \dd^3 \bx' a^{\dg \alpha} (\bx') (\beta m - i \alpha^i \p'_i)_{\alpha \beta}a^\beta(\bx'),
\lbl{4.67}
\ee
we have
$$
  [{\hat H},a_\delta^\dg (\bx)] =  \int \dd^3 \bx' \left [ a^{\dg \alpha} (\bx') (\beta m - i \alpha^i \p'_i)_{\alpha \beta}a^\beta (\bx') a^\dg_\delta (\bx)
  	-  a^\dg_\delta (\bx) a^{\dg \alpha} (\bx) (\beta m - i \alpha^i \p'_i)_{\beta \alpha}a^\beta(\bx')
  	\right ]$$
\be
  = (\beta m + i \alpha^i \p_i)_{\alpha \delta} a^{\alpha} (\bx)
  = h_{\alpha (\bx') \delta(\bx)} a^{\dg \alpha(\bx')} .
\lbl{4.68}
\ee
In the derivation above we used $a^{\dg \delta} (\bx) a^{\dg \alpha}(\bx') 
= -a^{\dg \alpha}(\bx') a^{\dg \delta} (\bx)$. 
The last equality in Eq.\,(\ref{4.68}) is true, because\footnote{
	For simplicity we omit the brackets, with understanding that the derivative $\p_i$ acts on
	the first term on its right side.}
$$
  h_{\beta(\bx')\alpha(\bx)} a^{\dg \beta(\bx')} = \int \dd^3 \bx'\, (\beta m - i \alpha^i \p'_i)_{\beta \al}
  \delta^3 (\bx'- \bx) a^{\dg \beta} (\bx')$$
\be 
  = \int \dd^3 \bx'\, (\beta m - i \alpha^i \p'_i)_{\beta \al}a^{\dg \beta} (\bx') \delta^3 (\bx'- \bx)
  = (\beta m + i \al^i \p_i)_{\beta \alpha} a^{\dg \beta} (\bx) .
\lbl{4.69}
\ee

The Heisenberg equation of motion (\ref{4.66}) for the operators $a^{\dg \al} (\bx)$ is
associated with the evolution equation $|\Psi (t) \rangle = e^{- i {\hat H} t} |\Psi (0) \rangle$.
Written in the form $ {\dot a}^{\dg \al} (\bx) = - i [{\hat H},a^{\dg \al} (\bx)]$, we see
that the sign in front of the commutator is opposite to the sign of the usual Heisenberg equation
for an operator ${\hat A}$, which are ${\dot {\hat A}} = i [{\hat H},{\hat A}]$, and follow from
$\langle \Psi (0)| {\rm e}^{i {\hat H} t} {\hat A}{\rm e}^{-i {\hat H} t}|\Psi(0) \rangle$.
Taking ${\hat A} = a^\al (t) A_{\al \beta} a^\beta (t)$ and 
$|\Psi(0) \rangle = \psi_\al (0) a^{\dg \al} (0) \vac$, we obtain
$$
 \langle 0| \psi^{*\rho} (0) a_\rho (0) {\rm e}^{i {\hat H}t} a^{\dg \al} (0) A_{\al \beta} a^\beta (0)
 {\rm e}^{-i {\hat H}t} a^{\dg \gam} (0) \psi_\gam (0) \vac$$
\be
  =\langle 0| \psi^{*\rho} (0) a_\rho (0) {\rm e}^{i {\hat H}t} a^{\dg \al} (0){\rm e}^{-i {\hat H}t} A_{\al \beta}{\rm e}^{i {\hat H}t} a^\beta (0)
  {\rm e}^{-i {\hat H}t} a^{\dg \gam} (0) \psi_\gam (0) \vac
\lbl{4.70}
\ee
Denoting
\be
  a_H^\dg (t) = {\rm e}^{i {\hat H}t} a^{\dg \al} (0){\rm e}^{-i {\hat H}t}~~~~~{\rm and}
  ~~~~~a_S^\dg (t) ={\rm e}^{-i {\hat H}t} a^{\dg \al} (0){\rm e}^{i {\hat H}t} ,
\lbl{4.71}
\ee
we distinguish two different kinds of evolution that starts from the initial operator
$a^{\dg \al} (0)$, $a^\al (0)$. The operators $a_S^{\dg \al} (t)$, $a_S^\al(t)$ determine the evolution of
a state $|\Psi (t)\rangle$ $= \psi_\al (0) a_S^{\dg \al} (t) \vac =$ 
${\rm e}^{-i {\hat H}t} |\Psi(0) \rangle$,
$\langle \Psi(t)| = \langle \Psi(0)|{\rm e}^{i {\hat H}t}$ $= \langle 0| a_S (t) \psi^* (0)$,
while the operators  $a_H^{\dg \al} (t)$, $a_H^\al(t)$ determine the evolution of an operator
${\hat A} (t) = a_H^{\dg \al} (t) A_{\al \beta} a_H^\beta (t) = $ 
${\rm e}^{i {\hat H}t} {\hat A} {\rm e}^{-i {\hat H}t}$.

\subsection{Multiparticle states}

Recalling now that $k_{i \al(\bx)}$, $k'_{i \al(\bx)} \equiv q_{i \al(\bx)}$ or
$a_{i \al(\bx)} \equiv (a_{\al (\bx)}^\dg,a_{\al (\bx)})$ are the generators of the
infinite-dimensional Clifford algebra $Cl(\infty)$. The space they span is a subset of $Cl(\infty))$.
The objects in the full space $C(\infty)$ can be written as
\be
  \Phi = \sum_{r=0}^\infty \frac{1}{\sqrt{r!}} \psi^{i_1 \al_1 (\bx_1) ...i_r \al_r (\bx_r)}
  a_{i_1 \al_1(\bx_1)}...a_{i_r \al_r (\bx_r)}.
\lbl{4.72}
\ee

In the case of fermions (the orthogonal case), we can define a vacuum state according to
\be
  \vac = \prod_{\bx,\al} a_\al (\bx).
\lbl{4.73}
\ee
The latter state is annihilated by any of the operators $a_\al (\bx)$, i.e., $a_\al (\bx) \vac = 0$.
Applying the state (\ref{4.72}) on the vacuum, we find
\be
 \Phi \vac = \sum_r \frac{1}{\sqrt{r!}} \psi^{\al_1 (\bx_1)
 	...\al_r (\bx_r)} a_{\al_1 (\bx_1)}^\dg...a_{\al_r (\bx_r)}^\dg
 \vac  .
\lbl{4.74}
\ee
This is a Fock state of the quantum field theory in the $\bx$-representation. This is not a usual state
considred in quantum field theory, bacause an operator
 $a_\al^\dg (\bx) \equiv {\hat \psi}_\al^\dg (\bx)$, expanded in terms of the momenta $\bp$, 
 creates positive and negative energy states which are considered a problematic.

In {\it momentum representation}, a possible vacuum is
\be
  \vac_B = \prod_{\sg,\bp} b_\sg (\bp) d_\sg^\dg  (\bp)~,~~~b_\sg (\bp) \vac_B = 0~, 
  d_\sg^\dg  (\bp) \vac_B = 0.
\lbl{4.75}
\ee
It is called {\it bare vacuum}. The basis Fock space one particle states are
\be
   b^\dg_\sg (\bp) \vac_B~,~~~ d_\sg (\bp) \vac_B ,
\lbl{4.76}
\ee
and their multiparticle generalization, where $b^\dg_\sg (\bp) \vac_B$ creates a positive energy
electron and $d_\sg (\bp)$ a negative energy electron.

In the Dirac theory, instead of $\vac_B$, one introduces the {\it Dirac vacuum}
\be
 \vac_D = \prod_{\sg,\bp} d_\sg (\bp) \vac_B ~,~~~~~  d_\sg (\bp) \vac_D = 0 .
\lbl{4.77}
\ee
Acting on $\vac_D$, the operators $d_\sg^\dg (\bp)$ create holes in the sea of negative energy
states, i.e., {\it antiparticles} (positrons). Particles (electrrons) are created by $b_\sg^\dg (\bp)$. 
The one particle states of the Dirac theory are thus
\be
b_\sg (\bp)^\dg \vac_D~,~~~ d_\sg^\dg (\bp) \vac_D .
\lbl{4.78}
\ee

It is often claimed that the concept of the Dirac vacuum as the sea of negative energy states is
nowadays obsolete and unnecessary. But from the point of view of orthogonal Clifford algebras and
fermionic fields as Clifford algebra valued objects, the concept of vacuum as the product of
operators is unavoidable\ci{BaezVacuum,PavsicOrthoSymp,PavsicFirence,PavsicStumbling} in
the finite-dimensional case, and thus naturally also occurs in infinite-dimensional case.

Within this scheme, the vacuum defined according to (\ref{4.73}) is just an idealization, a limiting 
case  of the Clifford algebra object of the sort
\be
  \frac{1}{\sqrt{r!}} \phi^{\al_1(\bx_1) \al_2 (\bx_2) ...\al_n (\bx_n)}
  a_{\al_1 (\bx_1)}....a_{\al_n (\bx_n)}  \longrightarrow \vac  
\lbl{4.79}
\ee
if $n \rightarrow \infty$ and for infinitely densely packed annihilation operators,
so that the set of $\bx_1,\bx_2,...,\bx_n$ approaches continuum. A realistic vacuum is thus of the form (\ref{4.79}). It is just one of the terms in the superposition (\ref{4.72}). The
component $\phi^{\al_1(\bx_1)...\al_n (\bx_n)}$ can be in principle any funcytion. In particular,
it can be a Gaussian function, its square being
$|\phi^{\al_1(\bx_1)...\al_n (\bx_n)}|^2 \propto {\rm exp}[\sum_{k=1}^n \frac{1}{\kappa_k} 
(\bx_k-{\bm X}_k)^2$.
It can also be of the form
\be
  |\phi^{\al_1(\bx_1)...\al_n (\bx_n)}|^2 \propto \prod_{k=1}^n \theta (\bx_k -{\bm a}_k)
  (1- \theta  (\bx_k -{\bm b}_k),
\lbl{4.80}
\ee
\setlength{\unitlength}{.68mm}
\begin{figure}[h!]
	
	\centerline{\includegraphics[scale=0.45]{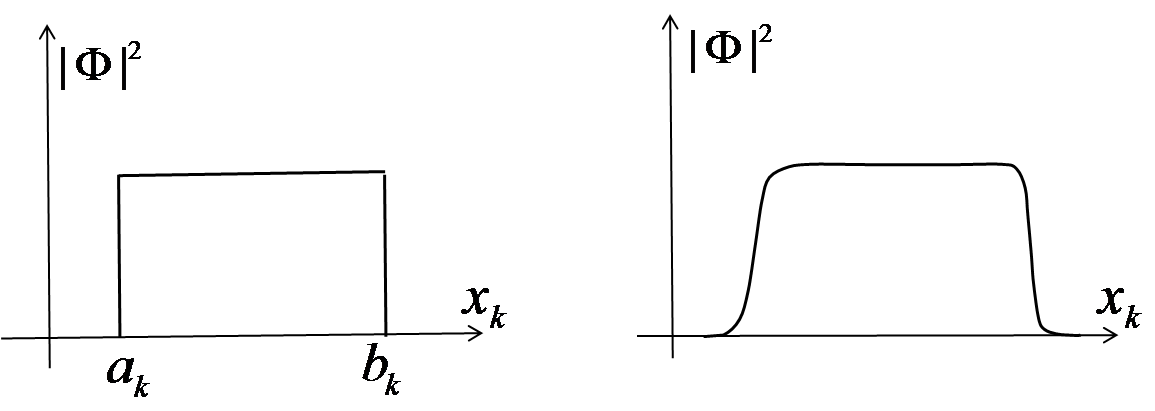}}
	
	\caption{\footnotesize An illustration of a physically realistic vacuum in position space with a sharp cutoff (left), and a smooth cutoff (right). The analogous figure also holds in momentum space.}
\end{figure}
i.e., having a constant finite value in the domain of positions between ${\bm a}_k$ and ${\bm b}_k$,
$k =1,2,...,n$ (illustrated in Fig.\,1-left). The sharp cutoffs implied by the function (\ref{4.80})
can be smoothed, as shown in Fig.\,1-right.  In general, the vacuum wave function can be anything between the
Gaussian wave function and the sharp cutoff wave function. Ina  spatially closed universe, the edges
${\bm a}_k$ and ${\bm b}_k$ can be identified, and the wave function is periodic, so that there are
no cutoffs.

In momentum representation, a generic state $\Phi$ is spanned over the basis states that consist
of the products of the generators
\be
  a_{i \sg(\bp)} = (b_{\sg(\bp)}^\dg, b_{\sg (\bp)},d_{\sg(\bp)},d_{\sg(\bp)}^\dg)~,~~~~i=1,2,3,4.
\lbl{4.81}
\ee
Thus,
$$
  \Phi = \sum_{r=1}^n \frac{1}{\sqrt{r!}}\phi^{i_1 \sg_1(\bp_1) i_2 \sg_2 (\bp_2)...i_r \sg_r (\bp_r)}
   a_{i_1 \sg_1(\bp_1)}a_{i_2 \sg_2 (\bp_2)}...a_{i_r \sg_r (\bp_r)}$$
$$
    = \Phi^{(0)} + \phi^{1 \sg(\bp)} b_{\sg(\bp)}^\dg + \frac{1}{\sqrt{2!}}\phi^{1 \sg_1 (\bp_1) 1 \sg_2 (\bp_2)}
     b_{\sg_1(\bp_1)}^\dg  b_{\sg_2(\bp_2)}^\dg+ ... $$
   $$  + \frac{1}{\sqrt{2!}}\phi^{2 \sg_1 (\bp_1) 4 \sg_2 (\bp_2)} b_{\sg_1(\bp_1)}  d_{\sg_2(\bp_2)}^\dg
      +	...$$
\bear
     + \frac{1}{\sqrt{n!}}\phi^{1 \sg_1 (\bp_1) 1 \sg_2 (\bp_2)...1 \sg_{n_1} (\bp_{n_1}) 4 \sg_1 (\bp_1) 4\sg_2 (\bp_2)
     ...4 \sg_{n_4} (\bp_{n_4})} b_{\sg_1(\bp_1)}...b_{\sg_{n_4}(\bp_{n_4})}
      d_{\sg_1(\bp_1)}^\dg ...d_{\sg_{n_4}(\bp_{b_4})}^\dg ~, \lbl{4.82}\\
      ~~n_1 + n_4 = n .\nonumber
\ear
The lat term in the above equation is a finite discrete version of the bare vacuum (\ref{4.75}). It is
multiplied by a wavefunction that in particular can be a Gaussian one,
\be
  |\phi^{1...1,4...4}|^2 \propto {\rm exp}\left [ \frac{1}{2 \kappa_1}(\bp_1 - \bP_1)^2 + 
   \frac{1}{2 \kappa_2}(\bp_2 - \bP_2)^2 + ...+  \frac{1}{2 \kappa_n}(\bp_n - \bP_n)^2 \right ] ,
\lbl{4.83}
\ee
or a step-like one,
\be
  |\phi^{1...1,4...4}|^2 \propto \prod_{k=1}^n \theta (\bp_k - \bP_k)
  (1- \theta  (\bp_k -\bP_k),
\lbl{4.84}
\ee
or its smoothed version, like in Fig.\,1-right.

If we multiply the $\Phi$ of Eq.\,(\ref{4.82}) with the vacuum
\be
  \vac_{n} = \frac{1}{\sqrt{n!}}\phi^{1 \sg_1 (\bp_1) ...1 \sg_{n_1} (\bp_{n_1}) 4 \sg_1 (\bp_1)
  	...4 \sg_{n_4} (\bp_{n_4})} b_{\sg_1(\bp_1)}...b_{\sg_{n_4}(\bp_{n_4})}
  d_{\sg_1(\bp_1)}^\dg ...d_{\sg_{n_4}(\bp_{b_4})}^\dg ,
\lbl{4.85}
\ee
we obtain the expression
\be
\Phi \vac_n =
\sum_{r_1=1}^{n_1} \sum_{r_2=1}^{n_2} 
 \frac{1}{\sqrt{n!}}\psi^{1 \sg_1 (\bp_1)... 4 \sg_1 (\bp_1)...}
  b_{\sg_1(\bp_1)}^\dg  b_{\sg_2(\bp_2)}^\dg...b_{\sg_{r_1}(\bp_{r_1})}^\dg 
   d_{\sg_1(\bp_1)} d_{\sg_2(\bp_2)}...d_{\sg_{r_2}(\bp_{r_2})}\vac_n ,
\lbl{4.86}
\ee
in which there are only the operatore $ b_{\sg(\bp)}^\dg$ and $d_{\sg(\bp)}$.
The wave functions $\psi^{1 \sg_1 (\bp_1)...
	4 \sg_1 (\bp_1)...}$ are linear combinations
of the wave functions $\phi^{i_1 \sg_1(\bp_1)...i_r \sg_r(\bp_r)}$, occurring in Eq.\,(\ref{4.82}), multiplied
by the vacuum wave function, occurring in Eq.\,(\ref{4.85}). If the wacuum wave function has
a cutoff (or a smoothed cutoff), then also the wave functions $\psi^{1 \sg_1 (\bp_1)..., 4 \sg_1 (\bp_1)...}$,
occurring in Eq.\,(\ref{4.86}) have a cutoff.

The state (\ref{4.86}) belongs to one of the many minimal left ideals of the Clifford algebra $C(\infty)$,
whose generic object is given by the expression (\ref{4.82}). The $\Phi \vac_n$ can thus be found within
the general expression (\ref{4.82}); it lives in a subspace of $C(\infty)$. The subspace, determined
by the vacuum $\vac_n$ of Eq.\,(\ref{4.85}) is a modification of the Fock space of the usual quantum field
theory for fermions, in which the bare vacuum is defined according to Eq.\,(\ref{4.75}). More precisely,
the vacuum of the usual QFT is not the bare vacuum, but the Dirac vacuum (\ref{4.77}), which is an
idealization. A realistic version of the Dirac vacuum is a state of the form
$$
  \vac_{nD} = \phi^{3 \sg_1 (\bp_1)...3 \sg_{n_3} (\bp_{n_3}) 1 \sg_1 (\bp_1)...1 \sg_{n_1} (\bp_{n_1})
  4 \sg_1 (\bp_1)...4 \sg_{n_4} (\bp_{n_4})}$$
\be
    \hs{2cm} \times d_{\sg_1 (\bp_1)}...d_{\sg_{n_3} (\bp_1)} b_{\sg_1 (\bp_1)}...b_{\sg_{n_1} (\bp_{n_1})}
    d_{\sg_1 (\bp_1)}^\dg...d_{\sg_{n_4} (\bp_{n_4})}^\dg ,
\lbl{4.87}
\ee
in which there can be a huge number of the states with negative energies, created by the operators 
$d_{\sg (\bp)}$.

\subsection{Antiparticles}

Acting on $\vac_{nD}$ with an operator $d_{\sg_k} (\bp_k)^\dg$, where $\bp_k \in \lbrace \bp_1,\bp_2,...,
bp_{n_3} \rbrace$, i.e., one of the states with the discrete momenta $\bp_1,\bp_2,...,\bp_{n_3}$, one creates a hole
in which the state (particle) with the quantum numbers $\sg_k$, $\bp_k$ is missing. Such holes are
{\it antiparticles}.

Recall from Sec.\,3 that the Dirac wave function $\psi^\al (t,\bx)$, $\al=1,2,3,4$ is the probability amplitude
of finding the particle at position $\bx$ at time $t$. The electric charge of a particle described by such
wave function is zero. Electric charge enters the game if one takes the pair of the Dirac wave functions
$(\psi_1^\al (\bx),\psi_2^\al (\bx))$. Then the pairs of annihilation operators $(a_{1 \al (\bx)},a_{2 \al (\bx)})$,
and their hermitian conjugates. In momentum representation we have the pairs $(b_{1 \sg (\bp)},b_{2 \sg (\bp)})$,
$(d_{1 \sg (\bp)}^\dg,d_{2 \sg (\bp)}^\dg)$ and the corresponding hermitian conjugates.
$(b_{1 \sg (\bp)}^\dg,b_{2 \sg (\bp)}^\sg)$, $(d_{1 \al (\bp)},d_{2 \al (\bp)})$. The states (\ref{4.81}) and,
in particular, (\ref{4.86}), have {\it zero electric charge}. The bare vacuum $\vac$ (Eq.\,\ref{4.75})),
the Dirac vacuum $\vac_D$ (Eq.\,\ref{4.77}), and their finite versions $\vac_N$, $\vac_{nD}$, then also have zero
electric charge. 

However, if we take the pairs of operators, then by interpreting
$b_{\sg (\bp)} \equiv (b_{1 \al (\bp)},b_{2 \al (\bp)})$,
$d_{\sg (\bp)}^\dg \equiv (d_{1 \al (\bp)}^\dg,d_{2 \al (\bp)}^\dg)$, etc., then Eqs.\,(\ref{4.81})--(\ref{4.87})
describe charged particles.

In analogy to \eq{4.49}), one can introduce
\be
  a_{+(\bx)} = \frac{1}{\sqrt{2}}(a_{1 (\bx)} + \bbi a_{2 (\bx)})~,~~~~
  a_{-(\bx)} = \frac{1}{\sqrt{2}}(a_{1 (\bx)} + \bbi a_{2 (\bx)}) ,
\lbl{4.88a}	
\ee
\be
b_{+\sg(\bp)} = \frac{1}{\sqrt{2}}(b_{1 \sg(\bp)} + \bbi a_{2 \sg(\bp)})~,~~~~
b_{-\sg(\bp)} = \frac{1}{\sqrt{2}}(b_{1 \sg(\bp)} - \bbi a_{2 \sg(\bp)})
\lbl{4.88ab}	
\ee
Thus, e.g., the operators $a_{+(\bx)}$ and $a_{+(\bx)}^\dg$, respectively, annihilate and create
a positively charged particle at position $\bx$. Similarly, $b_{+\sg(\bp)}$ and $b_{+\sg(\bp)}^\dg$,
respectively, annihilate and create a positively charged particle with momentum $\bp$ and spin
orientation $\sg=(1,2)$.

By defining
\be
  b_{\sg (\bp)} \equiv (b_{+\sg(\bp)},b_{-\sg(\bp)})~,~~~~
  d_{\sg (\bp)}^\dg \equiv (d_{+\sg(\bp)}^\dg,d_{-\sg(\bp)}^\dg)
\lbl{4.89}
\ee
\be
b_{\sg (\bp)}^\dg \equiv (b_{+\sg(\bp)}^\dg,b_{-\sg(\bp)}^\dg)~,~~~~
d_{\sg (\bp)} \equiv (d_{+\sg(\bp)},d_{-\sg(\bp)})
\lbl{4.90}
\ee
we can use the same equation (\ref{4.81})--(\ref{4.87}) for description of charged particles.
The Dirac vacuum $\vac_D$ (or its finite analog $\vac_{nD}$) is then a sea of equal quantity
(or number) of positively and negatively charged
particles, all with negative energies. Holes in that sea behave, respectively, as negatively and positively
charged particles, i.e., as antiparticles of the oppositely charged particles. Thus,
acting with the operators $d_{+\sg (\bp)}^\dg$ or $d_{-\sg (\bp)}^\dg$ on such Dirac vacuum one creates
holes with negative or positive charges. Acting on such vacuum with the operators
$b_{+\sg (\bp)}^\dg$ or $b_{-\sg (\bp)}^\dg$, one creates particles with positive or negative charges.
In this generalized quantum field theory, an electron can have a negative or positive electric charge. Such a
positively charged electron is not the antiparticle of the negatively charged electron. The respective
antiparticles of positively or negatively charged electrons are holes in the generalized Dirac vacuum, which
has zero electric charge. Such assignment of charges is only a toy model, an intermediate step in the development
of the theory, and will be modified within a more general framework, considered in Sec.,6, that includes
weak interaction.

\section{The presence of electromagnetic field}
  
Let us now include the electromagnetic field through minimal coupling, by replacing $\p/\p_\mu \equiv
\/\p x^\mu$ with the covariant derivative $\DD_\mu \equiv \p_\mu + \bbi A_\mu$. Instead of \eq{4.61}) and (\ref{4.66}) we have now the following equation
\be
  i \DD_0 \psi^{\beta (\bx')} = h_{\beta (\bx')\al(\bx)} \psi^{\al(\bx)} ,
\lbl{5.1}
\ee
\be
  i \DD_0 a^{\dg \beta (\bx')} = a^{\dg \beta(\bx')} h_{\beta(\bx') \al(\bx)} ,
\lbl{5.2}
\ee
where
\be
  h_{\al(\bx) \beta (\bx')} = \beta (m - i \gam^i \DD_i)_{\al \beta} \delta^3 (\bx-\bx'),
\lbl{5.3}
\ee
These are, respectively, the Dirac equation for a single-particle wave function $\psi^{\al(\bx)}$
and the field operators $a_{\al(\bx)} \equiv \psi_{\al(\bx)}$.

The action for the time dependent operators is
\be
  {\hat I} = \int \dd t \, \dd^3 \bx \,\left [ i a^{\dg \al} (\bx) \DD_0 a^\beta (\bx) \delta_{\al \beta}
  - a^{\dg \al} (\beta m - i\al^i \DD_i )_{\al \beta} a^\beta (\bx)\right ] .
\lbl{5.4}
\ee
The expectation value of the latter action in a single-particle state
\be
   |\Psi \rangle = \psi^{\al(\bx)} a_{\al(\bx)}^\dg \vac \equiv
   \int \dd^3 \bx \, \psi^\al (\bx) a_\al^\dg (\bx) \vac
\lbl{5.5}
\ee
is
$$
   I = \langle \Psi|{\hat I}|\Psi \rangle = \langle 0| \int \dd t\\, \dd^3 \bx \, \dd^3 \bx'\, \dd^3 \bx''\, 
    \psi^{* \gam} (\bx') a_\gam (\bx') \hs{6cm}$$
$$  
    \times \left [
   i a^{\dg \al)} (\bx) \DD_o a^\beta (\bx) \delta_{\al \beta} -
   	a^{\dg \al} (\bx) (\beta m - i \al^i \DD_i )_{\al \beta} a^\beta (\bx) \right ]
   	\psi^\delta (\bx'') a_\delta^\dg (\bx'') \vac
$$
\be
    = \langle 0| \int \dd t\, \dd^3 \bx \left [ i \psi^{* \beta} \DD_0 a_\beta \psi^\delta a_\delta^\dg
    - \psi^{* \al} (\beta m - i \al^i \DD_i )_{\al \beta} \psi^\beta \right ] \vac .
\lbl{5.6}
\ee
Let us now observe that
\be
  \langle 0| \int \dd t\, \dd^3 \bx \psi^{* \beta} \DD_0 a_\beta
  = \DD_0 \left ( \langle 0| \int \dd t\, \dd^3 \bx \psi^{* \beta}  a_\beta \right ) ,
\lbl{5.7}
\ee
where time dependence is on the operator. Alternatively, if time dependence is on the wave function,
then
\be
   \DD_0 \left ( \langle 0| \int \dd t\, \dd^3 \bx \psi^{* \beta}  a_\beta \right )
   = - \langle 0| \int \dd t\, \dd^3 \bx \DD_0 \psi^{* \beta}  a_\beta  
\lbl{5.8}
\end{equation}
Therefore, by switching time dependence from operators to to wave function, instead of \eq{5.6}),
we have
\be
  I = \langle \Psi|{\hat I}|\Psi \rangle = \int \dd t\, \dd^3 \bx \left [
  -i \DD_0 \psi^{* \al} \psi^\beta
   - \psi^{* \al} ( \beta m - i \al^i \DD_i )_{\al \beta} \psi^\beta \right ]   
\lbl{5.9}
\ee 
This is the action for a single-particle wave function in the presence of a background
electromagnetic field $A_\mu$.

We have seen that the single-particle states $|\Psi \rangle = \int \dd^3 \bx \psi^\al (\bx) a_\al^\dg (\bx) \vac$ are a part of quantum field theory, not only in the absence of interactions, but also in the presence of a background electromagnetic field (which itself arises from spinor quantum electrodynamics).
Their wave packet profiles $\psi^\al (\bx)$ satisfy the action principle \eq{5.9}), which is that of
relativistic quantum theory, in the particular case considered here, of the Dirac spinor theory, with
the following distinction. Namely, the gauge field $A_\mu$ arises from the non local gauge transformations
of the group $U(1)$, acting within the doublet $(\psi_1^\al , \psi_2^\al ) \equiv \psi^\al$.

In Sec.\,3 we obtained the free Dirac field by starting from the Klein-Gordon equation \eq{3.5}) for a
4-component {\it real} field $\vphi \equiv \vphi^\al (\bx)$. Let us now repeat the procedure by
taking a doublet of fiels $(\vphi_1^\al , \vphi_2^\al)$ that we represent as a complex field
\be
  \vphi^\al = \frac{1}{\sqrt{2}} (\vphi_1^\al + \bbi \vphi_2^\al)\equiv \vphi ~,~~~~
  \vphi^{{\bar *}\al} = \frac{1}{\sqrt{2}} (\vphi_1^\al - \bbi \vphi_2^\al) \equiv \vphi^{\bar *},
\lbl{5.10}
\ee
where the imaginary unit $\bbi$ is the generator of rotation between $\vphi_1^\al$ and $\vphi_2^\al$,
and ${\bar *}$ denotes complex conjugation with respect to $\bbi$. It satisfies the action
principle \eq{2.13}) or, equivalently, \eq{2.14}), this time for a 4-component complex field
$\vphi \equiv \vphi^\al$. By generalizing the action \eq{2.14}) so to become invariant under
local gauge transformations of the group $U_\bbi (1)$ of the transformations ${\rm e}^{\bbi \al(x)}$,
$x\equiv x^\mu$, we have
\be
  I[\vphi,\vphi^\bbs,\Pi,\Pi^\bbs]= \int \dd t \,\dd^3 \bx \left [ \Pi^\bbs \DD_0 \vphi +
  \Pi \DD_0 \vphi^\bbs - \left ( \Pi^\bbs \Pi + \vphi^\bbs h^2 \vphi \right ) \right ],
\lbl{5.11}
\ee
where
\be
h^2 = \DD_i \DD^i + m^2~, ~~~~ \DD_0 = \p_0 + \bbi e A_0~, ~~~~ \DD_i = \p_i + \bbi e A_i.
\lbl{5.12}
\ee
The square root of $h^2$ can be either (i) $h= \sqrt{m^2 + \DD_i \DD^i}$, 
(ii) $h = \beta m - i \al^i \p_i$, or, (iii) $h = \beta \sqrt{m^2 + \DD_i \DD^i}$.

Variation of the latter action with respect to ${\dot \vphi}^\bbs$ and ${\dot \vphi}$ gives
the cananocal momenta
\be
  \Pi = \frac{\p L}{\p {\dot \vphi}^\bbs} = \DD_0 \vphi = \p_0 \vphi + \bbi e A_0\vphi~,~~~~
  \Pi^\bbs = \frac{\p L}{\p {\dot \vphi}} = \DD_0 \vphi^\bbs = \p_0 \vphi^\bbs - \bbi \vphi^\bbs e A_0 .
\lbl{5.13}
\ee
Inserting the latter expression into the above phase space action, we obtain
\be
  I[\vphi,\vphi^\bbs] = \int \dd t \,\dd^3 \bx \left [ (\p_0 - \bbi A_0) \vphi^\bbs
  (\p_0 + \bbi A_0) \vphi + (\p_i - \bbi A_i) \vphi^\bbs (\p_i + \bbi A_i) \vphi
   - m^2 \vphi^\bbs \vphi \right ] ,
\lbl{5.14}
\ee
which is the Klein-Gordon equation in the preseence of a background electromagnetic field $A_\mu
= (A_0,A_i)$.

Introducing the new variables
\be
  \psi = \frac{1}{\sqrt{2}} \left ( \sqrt{h} \vphi + \frac{i}{\sqrt{h}} \Pi \right )~,~~~~
  \psi^* = \frac{1}{\sqrt{2}} \left ( \sqrt{h^*} \vphi + \frac{i}{\sqrt{h^*}} \Pi \right ) ,
\lbl{5.15}
\ee
\be
  {\tl \psi} = \frac{1}{\sqrt{2}} \left ( \sqrt{h} \vphi^\bbs + \frac{i}{\sqrt{h}} \Pi^\bbs \right )~,~~~~
\   {\tl \psi}^* = \frac{1}{\sqrt{2}} \left ( \sqrt{h^*} \vphi^\bbs
   + \frac{i}{\sqrt{h^*}} \Pi^\bbs \right ) ,
\lbl{5.16}
\ee
the action \eq{5.11}) reads
$$
  I[\psi,\psi^*,{\tl \psi},{\tl \psi}^\bbs] = \int \dd t\, \dd^3 \bx \left [ i {\tl \psi}^* \DD_0 \psi
  + i \psi^* \DD_0 {\tl \psi} - {\tl \psi}^* h \psi - \psi^* h {\tl \psi} \right ] \hs{3cm}$$
\be
  \hs{2cm} + ~ \text{terms involving commutators} ~~[\DD_0,\sqrt{h}]~~ {\rm and} ~~[\DD_0,\frac{1}{\sqrt{h}}].
\lbl{5.17}
\ee
This can be verified by plugging the expressions (\ref{5.15}) and (\ref{5.16}) into the action
(\ref{5.17}). For this aim it is useful to write the expressions (\ref{5.15}),(\ref{5.16}) in a
more explicite notation:
\be
  \psi^{\al(\bx)} = \left ( h_{\al(\bx)\beta(\bx')}^{1/2} \vphi^{\beta(\bx)} +
   i h_{\al(\bx)\beta(\bx')}^{-1/2} \Pi^{\beta(\bx)} \right )~,
\lbl{5.18}
\ee
\be
   {\tl \psi}^{\al(\bx)} = \left ( h_{\al(\bx)\beta(\bx')}^{1/2} \vphi^{\bbs \beta(\bx)} +
   i h_{\al(\bx)\beta(\bx')}^{-1/2} \Pi^{\bbs \beta(\bx)} \right ) 
\lbl{5.19}
\ee
and use the hermiticity of the operators $h$, namely, $h^\dg = h$, which in components
reads $h_{\al(\bx) \beta(\bx')}^\dg$ $= h_{\beta(\bx')\al(\bx)}^*$.
Expanding 
$$h^{1/2} = \sqrt{\beta m} \left ( 1 -  \frac{i\beta \al^i \p_i}{m} \right )^{1/2}
= \sqrt{\beta m} \left ( 1 - \frac{1}{2} \frac{i \beta \al^i \p_i}{m} 
- \frac{1}{2.4} \left ( \frac{i \beta \al^i \p_i}{m}  \right )^2 + ... \right )
$$
we find that $(h^*)^{1/2} = (h^{1/2})^*$ and $(h^\dg)^{1/2} = (h^{1/2})^\dg$, and the analogous for
$h^{-1/2}$.
Then we have
$$
 h_{\al(\bx)\beta(\bx')}^{*1/2} \vphi^{\beta(\bx')}
   = h_{\beta(\bx') \al(\bx)}^{\dg 1/2} \vphi^{\beta(\bx')}
  = h_{\beta(\bx') \al(\bx)}^{1/2} \vphi^{\beta(\bx')} \hs{3cm}$$
\be
  \hs{2cm} {\rm and} 
  ~~~~~~~ h_{\al(\bx)\beta(\bx')}^{* -1/2} \Pi^{\beta(\bx')}=h_{\beta(\bx') \al(\bx)}^{\dg -1/2} \Pi^{\beta(\bx')} \equiv \Pi h^{-1/2} .
\lbl{5.20}
\ee
Writing ${\tl \psi}^* \DD_0 \psi \equiv {\tl \psi}^{*\al} \DD_0 \psi^\beta \delta_{\al \beta}$
and using Eqs.\,\ref{5.18})--(\ref{5.20}) we find that the action (\ref{5.17}) assumes the
form (\ref{5.11}). In conformity with the usual notation of the Dirac equation we can
also write ${\tl \psi}^* \DD_0 \psi \equiv {\tl \psi}^\dg \DD_0 \psi$,
$\psi^* \DD_0 {\tl \psi} \equiv \psi^\dg \DD_0 {\tl \psi}$, ${\tl \psi}^* h \psi \equiv
{\tl \psi}^\dg h \psi$, and $\psi^* h {\tl \psi} \equiv \psi^\dg h {\tl \psi}$.

Varying the action (\ref{5.17}) with respect to ${\tl \psi}$, we obtain
\be
  i \DD_0 \psi = h \psi~ + ~ \text{\it extra terms with $[\DD_0,\sqrt{h}]$ and $[\DD_0,\frac{1}{\sqrt{h}}]$} .  
\lbl{5.21}
\ee
For the cases (i), (ii) and (iii) considered in \eq{5.12}) the latter equation gives, respectively,
\be
  (i) ~~~~i \p_0 \psi = \left (  i \bbi e A_0 + \sqrt{m^2 -\DD_i \DD^i} \right ) \psi ~ +
  \text{\it extra terms},
\lbl{5.22}
\ee
\be
(i) ~~~~i \p_0 \psi = \left (  i \bbi e A_0 + \beta m - i \al^i \p_i \right ) \psi ~ +
\text{\it extra terms},
\lbl{5.23}
\ee
\be
(i) ~~~~i \p_0 \psi = \left (  i \bbi e A_0 + \sqrt{m^2 -\DD_i \DD^i} \beta \right )  \psi ~ +
\text{\it extra terms},
\lbl{5.24}
\ee
where ``{\it extra terms''} contain the commutators $[\DD_0,\sqrt{h}]$ and $[\DD_0,\frac{1}{\sqrt{h}}]$.

Recall that we distinguish two different imaginary units:
\begin{quote}
  	$i$ ~ rotates~~~ $ \vphi \rightleftarrows \Pi~,~~~~ \vphi^\bbs \rightleftarrows \Pi^\bbs$\\
    $\bbi$ ~ rotates~~~ $\vphi \rightleftarrows \vphi_2~,~~~~ \Pi_1 \rightleftarrows \Pi_2$.
\end{quote}

The action (\ref{5.17}), obtained from the action action (\ref{5.11}) contains the extra terms,
because the commutators $[\DD_0,\sqrt{h}]$ and $[\DD_0,\frac{1}{\sqrt{h}}]$ do not vanish.
They vanish only in free case in which $\DD_0 = \p_0$. Therefore, if we consider the scalar
field which is minimally coupled to the electromagnetic field according to the action (\ref{5.11}),
then we find that such theory does not lead to the usual Dirac theory and the Dirac equation in the presence of a background
electromagnetic field $A_\mu = (A_0,A_i)$, $i = 1,2,3$. We must proceed in the opposite direction,
namely, to assume ---as usually---  the validity of the Dirac equation,
\be
  i \DD_0 \psi = \left ( \beta m - i \al^i \DD_i \right ) \psi ,
\lbl{5.24a}
\ee
in which $\psi$ is minimally coupled to $A_\mu$, and find out the corresponding second order equations.

Rewriting \eq{5.24a}) in the form
\be
  i \gam^0 \DD_0 \psi = (- i \gam^i \DD_i + m) \psi ,
\lbl{5.25}
\ee
and multiplying it with $i \gam^0 \DD_0$, we obtain
$$
   -\DD_0^2 \psi =  i \gam^0 \DD_0 (- i \gam^i \DD_i + m) \psi  \hs{4cm} $$
$$
    = \left ( \gam^0 \gam^i [\DD_0,\DD_i ] + (i \gam^i \DD_i + m) i \gam^0 \DD_0 \right ) \psi$$
\be
  = \left ( \gam^0 \gam^i [\DD_0,\DD_i ] + \gam^i \gam^j \DD_i \DD_j + m^2 \right ) \psi ,
\lbl{5.26}
\ee
i.e.,
\be
  \left ( \DD_0^2 + \gam^0 \gam^i [\DD_0,\DD_i ] + \gam^i \gam^j \DD_i \DD_j + m^2 \right ) \psi = 0 ,
\lbl{5.27}
\ee
or,
\be 
  (\gam^\mu \gam^\nu \DD_\mu \DD_\nu + m^2 ) \psi =0~, 
  ~~~{\rm where} ~~~[ \DD_\mu,\DD_\nu] = \bbi e F_{\mu \nu},
\lbl{5.27a}
\ee
which is a well known result, appart of the $\bbi$ instead of $i$.

The Dirac field satisfies the minimal action principle
\be
  I[\psi,\psi^\dg] = \int \dd^4 x \left ( i \psi^\dg \DD_0 \psi - \psi^\dg h \psi \right ).
\lbl{5.28}
\ee
Usually, it is taken that $\psi \equiv \psi^\al$ is a 4-component complex field. In this new
approach we take $\psi = (\psi_1^\al,\psi_2^\al)$, namely, an object that consists of two
4-comppnent complex fields $\psi_1^\al$, $\psi_2^\al$, satisfying the action principle
\be
  I[\psi_1,\psi_2,\psi_1^\dg,\psi_2^\dg] = \int \dd t\,\dd^3 \bx \, \left (
  \psi_1^\dg \left ( i \DD_0  -  h \right ) \psi_1
   + \psi_2^\dg \left ( i \DD_0  -  h \right ) \psi_2 \right ) .
\lbl{5.29}
\ee
In terms of the new variables  (\ref{2.48}), the action (\ref{5.29}) becomes
\be
  I[\chi_+,\chi_-, \chi_+^\dg,\chi_-^\dg] = \int \dd t \,\dd^3 \bx 
  \left ( \chi_+^\dg ( i \DD_0 - h ) \chi_- + \chi_-^\dg (i \DD_0 - h ) \chi_+\right ) ,
\lbl{5.30}
\ee
where $\chi_+$ and $\chi_-$ are wave functions bearing positive and negative charge,
respectively.

Variation of the action (\ref{5.30}) with respect to $\chi_+^\dg$ and $\chi_-^\dg$ gives
\be
  i \DD_0 \chi_+ = h \chi_+ ,
\lbl{5.31}
\ee
\be
i \DD_0 \chi_- = h \chi_- ,
\lbl{5.32}
\ee
i.e.,
\be
  i \DD_0 \begin{pmatrix}
  	  \chi_+\\
  	  \chi_-
    \end{pmatrix} 
  = h \begin{pmatrix}
  	\chi_+\\
  	\chi_-
  \end{pmatrix} ,
\lbl{5.33}
\ee
or shortly,
\be
  i \DD_0 \psi = h \psi~,~~~~ 
  \psi = \begin{pmatrix}
  	\chi_+\\
  	\chi_-
  \end{pmatrix} .
\lbl{5.34}
\ee

Let us rewrite the action (\ref{5.30}) in terms of two 4-component real field $\vphi_1$,
$\vphi_2$, related to $\psi_1$ and $\psi_2$ according to \eq{3.43}). Introducing
\be
  \vphi_\pm = \sqr ( \vphi_1 \pm \bbi \vphi_2 )~,~~~~~
  \vphi_\pm^\dg = \sqr ( \vphi_1^\dg \pm \bbi \vphi_2^\dg ) ,
\lbl{5.34a}
\ee
which in turn is related to  $\chi_+^\dg$ and $\chi_-^\dg$ according to
\be
  \chi_\pm = \sqr \left ( \sqrt{h}\vphi_\pm + \bbi\, \frac{1}{\sqrt{h}} \Pi_\pm \right )~,~~~~~
  \chi_\pm^\dg = \sqr \left (\vphi_\pm  \sqrt{h} - \bbi \, \Pi_\pm \frac{1}{\sqrt{h}}\right ) ,
\lbl{5.35}
\ee
where $(\sqrt{h})^\dg = \sqrt{h}$ and $\frac{1}{(\sqrt{h})^\dg} = \frac{1}{\sqrt{h}}$, as a consequence
of $h^\dg = h$.

Inserting (\ref{5.35}) into the action (\ref{5.30}), we obtain
$$
  I[\vphi_\pm,\Pi_\pm,\vphi_\pm^\dg,\Pi_\pm^\dg] = \int \dd t\, \dd^3\bx \frac{1}{2}
  \left [ \vphi_+^\dg \sqrt{h} K \sqrt{h} \vphi_- - i \Pi_+^\dg \frac{1}{\sqrt{h}} K \sqrt{h} \vphi_-
  \right .\hs{2cm} $$
 $$\hs{2cm} + i \vphi_+^\dg \sqrt{h} K \frac{1}{\sqrt{h}} \Pi_-
  + \Pi_+^\dg \frac{1}{\sqrt{h}} K \frac{1}{\sqrt{h}} \Pi_-$$
 $$ \hs{2cm}+ \vphi_-^\dg \sqrt{h} K \sqrt{h} \vphi_+ - i \Pi_- \frac{1}{\sqrt{h}} K \sqrt{h} \vphi_+ $$
\be
   \hs{2cm} \left . + i \vphi_-^\dg \sqrt{h} K \frac{1}{\sqrt{h}} \Pi_+
  + \Pi_-^\dg \frac{1}{\sqrt{h}} K \frac{1}{\sqrt{h}} \Pi_+ \right ] ,
\lbl{5.36}
\ee
where $K = i \DD_0 - h$, $h= \beta m - i \al^i \DD_i$. This is equal to
$$
  I[\vphi_\pm,\Pi_\pm,\vphi_\pm^\dg,\Pi_\pm^\dg] = \int \dd t\, \dd^3\bx \left [
  \Pi_+^\dg \DD_0 \vphi_- +  \Pi_-^\dg \DD_0 \vphi_+ - (\Pi_+^\dg \Pi_- + \vphi_+^\dg h^2 \vphi \vphi_- )
  \right . \hs{2cm}$$
\be
  \hs{2cm}+ \left .  \Pi_+ \frac{1}{\sqrt{h}} [\DD_0,\sqrt{h}] \vphi_- + ...
  	+ \vphi_+^\dg \sqrt{h} \DD_0 \sqrt{h} \vphi_- 
  	+ \Pi_+^\dg \frac{1}{\sqrt{h}} \DD_0 \frac{1}{\sqrt{h}} \Pi_- + ... \right ]  ,
\lbl{5.37}
\ee
where ``...'' denotes the analogous terms,

Variation of the action (\ref{5.36}) with respect to $\vphi_\pm^\dg$ gives
\be
  \sqrt{h} K \sqrt{h} \vphi_\mp + i \sqrt{h} K \frac{1}{\sqrt{h}} \Pi_\mp = 0,
\lbl{5.38}
\ee
or explicitly,
\be
  i \sqrt{h} [\DD_0,\sqrt{h}] \vphi_\mp + i h \DD_0 \vphi_\mp - h^2 \vphi_\mp - 
  \sqrt{h} [\DD_0,\frac{1}{\sqrt{h}}] \Pi_\mp - \DD_0 \Pi_\mp - i h \Pi_\mp = 0 .
\lbl{5.39}
\ee
Using
\be
  [\DD_0,\sqrt{h}] = [\DD_0, \frac{1}{\sqrt{h}} h ] = \frac{1}{\sqrt{h}} [\DD_0,h] +
  [\DD_0,\frac{1}{\sqrt{h}}] h ,
\lbl{5.40}
\ee
equation (\ref{5.39}) becomes
\be
  \DD_0 \Pi_\mp + h^2 \vphi_\mp - i [\DD_0,h] \vphi_\mp
   - i \sqrt{h} [\DD_0,\frac{1}{\sqrt{h}}] (h \vphi_\mp + i \Pi_\mp )
   + i h (\Pi_\mp - \DD_0 \vphi_\mp) = 0 .
\lbl{5.41}
\ee
Here
\be
  - i [\DD_0,h] = \gam^0 \gam^i \bbi e F_{0i} ~,~~~~ \bbi e F_{0i} = [\DD_0,\DD_i] ,
\lbl{5.42}
\ee
and
\be
  h^2 = m^2 + \gam^i \gam^j \DD_i \DD_j = m^2 + \eta^{ij} \DD_i \DD_j +
  \frac{1}{4} [\gam^i,\gam^j] \bbi e F_{ij} ~,~~~~ \bbi e F_{ij} = [\DD_I,\DD_j] .
\lbl{5.43}
\ee
 
Equation (\ref{5.41}) contains real and imaginary part and hence two equations for
the variables $\vphi_\pm$ and $\Pi_\pm$. However, to disentangle the real and imaginary
part is not straightforward, because of the occurrenceof the $i$ in $\sqrt{h}$ and
$\frac{1}{\sqrt{h}}$. As a first ansatz we can take the relation $\Pi_\pm = \DD_0 \vphi_\pm$.
Then, as the first iteration, we obtain
\be
   \DD_0^2 \vphi_\pm + h^2 \vphi_\pm - i [\DD_0,h] \vphi_\pm
  - i \sqrt{h} [\DD_0,\frac{1}{\sqrt{h}}] (h \vphi_\pm + i \DD_0 \vphi_\pm ) = 0  .
\lbl{5.44}
\ee

The same equation (\ref{5.41}) we obtain if we take the Dirac equation (\ref{5.34}) and
insert into it the expression (\ref{5.35}) for $\chi_\pm$. But if instead of the variables
$\vphi_\pm$ and $\Pi_\pm$ we take the new variables $\vphi'_\pm = \vphi/\sqrt{h}$ and
$\Pi'_\pm = \Pi_\pm/\sqrt{h}$, so that
\be
  \chi_\pm = \sqr ( h \vphi'_\pm + i \Pi'_\pm ) = \sqrt{\frac{h}{2}} \left (
  \sqrt{h} \vphi'_\pm + \frac{i}{\sqrt{h}} \Pi'_\pm \right ),
\lbl{5.45}
\ee 
then \eq{5.34}) gives
\be
  i [\DD_0,h] \vphi'_\pm + i h \DD_0 \vphi'_\pm - \DD_0 \Pi'_\pm = h^2 \vphi'_\pm + i h \Pi'_\pm
\lbl{5.46}
\ee
Taking into account that $-i [\DD_0,h] \vphi'_\pm = \gam^0 \gam^i \bbi e F_{0i}$ and
$h^2 = m^2 + \gam^i \gam^j \DD_i \DD_j$ are real\footnote{
	This is so if we consider $\gam_\mu$, $i$, and $\bbi$ as abstract objects, spanning the bases
	of different spaces. Namely, in geometric algebra, $\gam_\mu$ are Clifford algebra-valued objects and not matrices. Matrices are their representation on a spinor basis.},
we find that the imaginary part of the equation is
\be
  i \beta m \DD_0 \vphi'_\pm = i \beta m \Pi'_\pm ,
\lbl{5.47}
\ee
from which we have $\Pi'_\pm = \DD_0 \vphi'_\pm$. Inserting this into \eq{5.46}), we obtain
\be
  \DD_0^2 \vphi'_\pm + {\tilde h}^2 \vphi'_\pm = 0 ,
\lbl{5.48}
\ee
where
\be
  {\tl h}^2 = h^2 - i [\DD_0,h] = m^2 + \eta^{ij} \DD_i \DD_j
   + \frac{1}{4} [\gam^i,\gam^j] \bbi e F_{ij} + \gam^0 \gam^i \bbi e F_{0i} .
\lbl{5.49}
\ee
A compact form of \eq{5.48}) is
\be
  \left ( \gam^\mu \gam^\nu \DD_\mu \DD_nu + m^2 \right ) \vphi'_\pm = 0.
\lbl{5.50}
\ee

We have thus found that the Dirac equation for a {\it complex} valued pair of the field
$\chi_+$ and $\chi_-$ can be rewritten as a second order equation for the pair of the
{\it real} fields $\vphi'_+$, $\vphi'_-$, which are related to $\chi_+$, $\chi_-$
according to \eq{5.46}). The fields   $\vphi'_+$ and $\vphi'_-$ are real with respect to
the imaginary unit $i$, but they are complex with respect to the imaginary unit $\bbi$.

We can now use the relation $\Pi'_\pm = \DD_0 \vphi'_\pm$ which in terms of the old variables
reads
\be
  \Pi'_\pm = \sqrt{h} \DD_o \left ( \frac{1}{\sqrt{h}} \vphi_\pm \right )
   = \DD_0 \vphi_\pm +\sqrt{h} [\DD_0,\frac{1}{\sqrt{h}}] \vphi_\pm ,
\lbl{5.51}
\ee
and insert it into Eq.\,(\ref{5.41}). So we obtain
\be
  \DD_0^2 \vphi_\pm + {\tl h}^2 \vphi_\pm + {\cal O} = 0 ,
\lbl{5.52}
\ee
where ${\cal O}$ denotes the terms containing $[\DD_0,\sqrt{h}]$ and $[\DD_0,\frac{1}{\sqrt{h}}]$.

In terms of the new variables,
\be
  \chi'_\pm = \frac{1}{\sqrt{2}} \left ( h \beta \vphi'_\pm + i \DD_0 \vphi'_\pm \right ) ,
\lbl{5.53}
\ee
the equation (\ref{5.48}) assumes the following first order form:
\be
  i \DD_0 \chi'_\pm = {\tl h} \chi'_\pm + \left ( i [\DD_0,{\tl h} \beta] 
  - {\tl h} \beta [{\tl h},\beta] \right ) \frac{1}{\sqrt{2}} \left ( \chi^\dg \frac{1}{\tl h} + 
  \frac{1}{\tl h} \chi \right ) .
\lbl{5.54}
\ee
In free case $[\DD_0,{\tl h}]=0$, and the latter equation reads
\be
   i \DD_0 \chi'_\pm =\beta \sqrt{m^2 - \p^i \p_i }\, \chi'_\pm .
\lbl{5.55}
\ee
This is the Foldy-Wouthuysen transform of the free Dirac equation in which positive
energy components of the wave function are decoupled from those with negative energies, discussed
in Sec.\,3.

\section{Prospects and further clarifications}

\subsection{Extending to $U(1)\times SU(2)$}

In order to include the weak interaction, we have to extend the model. One possibility is to
double the wave functions $(\chi_+,\chi_-)$ and the corresponding creation operators $(b_+^\dg,b_-^\dg )$,
and consider the objects
\be
  \chi = \begin{pmatrix} \chi_{+1}\\ \chi_{+2}\\ \chi_{-2}\\ \chi_{-1}
   \end{pmatrix}
  ~~~{\rm and}~~~~ b^\dg = \begin{pmatrix} b_{+1}^\dg\\ b_{+2}^\dg\\ b_{-2}^\dg\\ b_{-1}^\dg
  \end{pmatrix}
\lbl{6.1}
\ee
Here $\chi_{+1} = \frac{1}{\sqrt{2}} (\psi_{11} - \bbi \psi_{12})$,
$b_{+1}^\dg = \frac{1}{\sqrt{2}} (b_{11}^\dg + \bbi b_{12}^\dg )$, and the analogous for
$\chi_{+2}$, $\chi_{-1}$, $\chi_{-2}$, $b_{+2}^\dg$, $b_{-1}^\dg$, $b_{-2}^\dg$. Each of those
components form a representation of the $U(1)$ whose elements are of the form
${\rm e}^{i \alpha {\hat y}/2}$. The doublets $b_+ = \begin{pmatrix} b_{+1}\\ b_{+2} \end{pmatrix}$ and $b_- = \begin{pmatrix} b_{-2}\\ b_{-1} \end{pmatrix}$ form two distinct
representations of of $SU(2)$.

There can be the hypercharge, weak isospin and charge quantum numbers asshown in Table 1.
\begin{center}
	Table 1: Assignemnt of quantum numbers to four types of  leptons\\
	\vs{2mm}
	\begin{tabular}{|c |c| c c c|} 
		\hline
		&~&  $~\,\frac{y}{2}$ & \,\,\,$I_3$ & ~~Q \\[0.5ex]
		\hline
		$~~b_{+1}^\dg$ &~~$\epsilon^+$ & $~~\,\frac{1}{2}$ & $~~\,\frac{1}{2}$&~~ 1 \\ [0.5ex]
		\hline
		$~~b_{+2}^\dg$&~~$\nu_\epsilon$ & ~~\,$\frac{1}{2}$ & ~~-$\frac{1}{2}$ &~~ 0 \\ [0.5ex]
		\hline
		$~~b_{-2}^\dg$&$~~\nu_e$ & ~~-$\frac{1}{2}$& ~~\,$\frac{1}{2}$ & ~~$~0$ \\ [0.5ex]
		\hline
		$~~b_{-1}^\dg$&$~~e^-$& ~~-$\frac{1}{2}$& ~~-$\frac{1}{2}$ & ~\,\,-$1$\\ [0.5ex]
		\hline
	\end{tabular}
\end{center}

Besides those particle states there are the corresponding antiparticle states
$\epsilon^-$, ${\bar \nu}_\epsilon$, ${\bar \nu}_e$, $e^+$. They are the holes
in the Dirac sea of negative energy states, created by $d_+ = (d_{+1},d_{+2})$ and
$d_- = (d_{-1},d_{-2})$.

One of possible scenarios is that both doublets, $(\nu_e,e^-)$ and $(\epsilon^+,\nu_\epsilon)$, are
coupled to the same gauge field of the local $U(1)\times SU(2)$ and thus feel the same electroweak
force.
In another possible scenario, the doublets $(\nu_e,e^-)$ and $(\epsilon^+,\nu_\epsilon)$
are coupled to different gauge fields associated with two different copies of $U(1) \times SU(2)$.
The particles $\epsilon^+$, $\nu_\epsilon$ are then invisible to the particles $e^-$ and $\nu_e$.

The same also holds for the quarks. They also occur in this model in two kinds of
$SU(2)$ doublets. Besides $(u,d)$, which is analogous to $(\nu_e, e^-)$, we also have
another doublet, $(u',d')$, which is analogous to $(\epsilon^+,\nu_\epsilon)$. According to the first scenario,
those two kinds of quark doublets are coupled to the same gauge field, while according to the second scenario
they are coupled to two different kinds of the $U(1)\times SU(2)$ gauge
fields and therefore invisible to each other. Then, the matter composed of $u'$, $d'$, $\epsilon^+$,
$\nu_\epsilon$ does not interact via an $U(1)\times SU(2)$ gauge fields with the matter
composed of $u$, $d$, $e^-$, $\nu_e$ (i.e., the matter we are composed of). For us it is
thus ``dark matter''. Such a model thus incorporates the notorious {\it dark matter}, detected via
gravitational effects in astronomical observations.

To justify the doubling of the wave function $\chi_+ = \frac{1}{2} (\psi_1 + \bbi \psi_2)$, let us
observe that the imaginary unit $\bbi$ (which is different from the usual $i$ can be considered
as one of the three quaternionic imaginary units ${\bm i}$,${\bm j}$,$\bm k$. The wave function
$\chi_+$ is then just a special case of a generic quaternion
\be
 \hs{2cm} {\bm \Phi} = {\tl \phi}^0 {\bm 1}+ {\tl \phi}^1 {\bm i} + {\tl \phi}^2 {\bm j} + {\tl \phi}^3 {\bm k}~,
  ~~~~ {\bm i} {\bm j} {\bm k} = -1~,~~ {\bm i \bm j} = {\bm k} ,
\lbl{6.2}
\ee  
which can be written as
\be
  {\bm \Phi}=  ( {\tl \phi}^0 {\bm 1}+ {\tl \phi}^1 {\bm i} ) {\bm 1}
  + ( {\tl \phi}^2 {\bm 1}+ {\tl \phi}^3 {\bm i}) {\bm k} .
\lbl{6.3}
\ee
Identifying
\be
  {\bm i} \equiv \bbi~,~~{\tl \phi}^0 = \frac{1}{\sqrt{2}} \psi_{11}~,~~{\tl \phi}^1 = \frac{1}{\sqrt{2}} \psi_{12}~,~~
{\tl \phi}^2 = \frac{1}{\sqrt{2}} \psi_{21}~,~~{\tl \phi}^3 = \frac{1}{\sqrt{2}} \psi_{22},
\lbl{6.3a}
\ee
we can write Eq.\,(\ref{6.3}) as
\be
  {\bm \Phi} = \chi_{+1} {\bm 1} + \chi_{+2} {\bm k}.
\lbl{6.4}
\ee

A generic ${\bm \Phi}$ can be considered as an element of $Cl(2) \otimes {\mathbb{C}}$, i.e., of a complexified
Clifford algebra in 2-dimensions. Namely, using
\be
  {\bm i} = i e_1~,~~{\bm j}= i e_2~,~~{\bm k} = {\bm i}{\bm j} = - e_1 e_2 ,
\lbl{6.5}
\ee
and redefining the coefficients in Eq.\,(\ref{6.2}) according to
\be
  \phi^0 = {\tl \phi}^0~,~~\phi^1 = i {\tl \phi}^1~,~~, \phi^2 = i {\tl \phi}^2~,~~\phi^{12} = - {\tl \phi}^3 ,
\lbl{6.5a}
\ee
we obtain
\be
  {\bm \Phi} = \phi^0 {\bm 1} + \phi^1 e_1 + \Phi^2 e_2 + \phi^{12} e_1 e_2 ,
\lbl{6.6}
\ee
where $e_1$, $e_2$ are generators of $Cl(2)$.

In terms of the spinor basis,
\be
  u_1 = \half (1+e_1)~,~~~u_2 = \half (e_2 - e_1 e_2)~,~~~
  u_3 = \half (1-e_1)~,~~~u_4 = \half (e_2 + e_1 e_2)~,
\lbl{6.7}
\ee
the Clifford algebra valued wave function $\bm \Phi$ assumes the following form:
$$
  {\bm \Phi} = \sqr \left [ (\psi_{11}+i \psi_{12}) u_1 + (\psi_{21}-i \psi_{22}) u_2 +
  (\psi_{21}+i \psi_{22}) u_3 + (\psi_{11}-i \psi_{12}) u_4 \right ]
$$
\be
  = {\tl \psi}^1 u_1 + {\tl \psi}^2 u_2 + {\tl \psi}^3 u_3 + {\tl \psi}^4 u_4 \hs{5cm},
\lbl{6.8}
\ee
where we have used the relations (\ref{6.3a}) and (\ref{6.5a}). Here $(u_1,u_2)$ ad $(u_3,u_4)$ span
two independent left ideals of $Cl(2)$, i.e., two independent spinors.

The following matrix representation of the $Cl(2)$ basis exists:
\be
  {\bm 1}= \begin{pmatrix} 1 & 0\\ 0 & 1
  \end{pmatrix}~,~~~
   e_1= \begin{pmatrix} 1 & 0\\ 0 & -1
   \end{pmatrix}~,~~~
   e_2= \begin{pmatrix} 0 & 1\\ 1 & 0
   \end{pmatrix}~,~~~
   e_1 e_2= \begin{pmatrix} 0 & 1\\ -1 & 0
   \end{pmatrix}~,
\lbl{6.9}
\ee
from which we have the relation
\be
  e_1 = \sg_3~,~~~e_2 = \sg_1~,~~~e_1 e_2 = i \sg_2 .
\lbl{6.9a}
\ee
where $\sg_1$, $\sg_2$, $\sg_3$ are the Pauli matrices.

The matrix representation of the basis spinors is then
\be
   u_1= \begin{pmatrix} 1 & 0\\ 0 & 0
   \end{pmatrix}~,~~~
   u_2= \begin{pmatrix} 0 & 0\\ 1 & 0
   \end{pmatrix}~,~~~
   u_3= \begin{pmatrix} 0 & 1\\ 0 & 0
   \end{pmatrix}~,~~~
   u_4= \begin{pmatrix} 0 & 0\\ 0 & 1
   \end{pmatrix} .
\lbl{6.10}
\ee

The basis spinors, $u_1,u_2$ of the first {\it left} ideal are the eigenstates of $e_1$,
and so are the basis spinors $u_3,u_4$ of the second {\it left} ideal:
\bear
  &&e_1 u_1 = u_1~,~~~~~~~ e_1 u_2 =- u_2 , \lbl{6.11}\\
  && e_1 u_3 = -u_3~,~~~~~ e_1 u_4 = u_4 \, .
\lbl{6.12}
\ear

The pairs $(u_1,u_2)$ and $(u_3,u_4)$ form two distinct representations of $SU(2)$.
An element ${\rm e}^{i \sg_a \al^a/2} \in SU(2)$, $a=1,2,3$, can act on the object
$\bm \psi$ from the left. Then it separately reshuffles the components of $u_1$ and $u_2$ on the one hand
and the components of $u_3$ and $u_4$ on the other hand. So we have, e.g.,
\be
   {\rm e}^{\frac{i \sg_2 \al}{2}} u_1 = \left ( {\rm cos} \, \frac{\al}{2}
   + e_1 e_2\, {\rm sin}\, \frac{\al}{2} \right )
   	\frac{1+e_1}{2} =  u_1 \,{\rm cos} \, \frac{\al}{2} +  u_2 \,{\rm sin}\, \frac{\al}{2}  ,
\lbl{6.13}
\ee
\be
   {\rm e}^{\frac{i \sg_2 \al}{2}} u_3 = \left ( {\rm cos} \, \frac{\al}{2}
   + e_1 e_2\,{\rm sin}\, \frac{\al}{2} \right )
	\frac{e_2 + e_1 e_2}{2} = u_3\,{\rm cos} \, \frac{\al}{2}  - u_4 \,{\rm sin}\, \frac{\al}{2}  ,
\lbl{6.14}
\ee
where we have used Eqs.\,(\ref{6.9a}) and (\ref{6.7}).
If acting from the right, then the transformation ${\rm e}^{i \sg_a \al^a/2}$ reshuffles the
components of $u_1$, $u_3$ within the first {\it right} ideal and the components $u_2$, $u_4$ within
the second {\it right} ideal.

Each of the wave functions in Eq.\,(\ref{6.8}) is a 4-component spinor, hence $\bm \Phi \equiv {\bm\Phi}^{\al (\bx)}$, $\al = 1,2,3,4$. By also taking into account the correspondin 4-spinor basis, given in terms
of the creation operators $a^\dg \equiv a_{\al(\bx)}^\dg$ or $b^\dg \equiv b_{\al(\bp)}^\dg$, we have the following
state:
\be
   {\bm \Phi} a^\dg \vac \equiv {\bm \Phi}^{\al(\bx)} a_{\al(\bx)} \vac
    = {\bm \Phi}^{\al(\bp)} b_{\al(\bp)} \vac.
\lbl{6.15}
\ee
The basis vectors are then $a^\dg u_1$,  $a^\dg u_2$,  $a^\dg u_3$,  $a^\dg u_4$,

Taking into account $e_1 = - i \bbi$ (see Eqs.\,(\ref{6.3a}), (\ref{6.5})) and defining $a_1 = \sqr a^\dg$,
 $a_2^\dg = - \frac{i }{\sqrt{2}}a^\dg$,
we have
\be
  a^\dg u_1 = \sqr (a_1^\dg + \bbi a_2^\dg) = a_+^\dg \equiv a_{+1}^\dg~.~~~~~ 
  a^\dg u_4 = \sqr (a_1^\dg - \bbi a_2^\dg) = a_-^\dg \equiv a_{-1}^\dg
\lbl{6.16}
\ee
\be
a^\dg u_2 = e_2 \sqr (a_1^\dg + \bbi a_2^\dg) = e_2 a_+^\dg \equiv a_{+2}^\dg~.~~~~~ 
a^\dg u_3 = e_2 \sqr (a_1^\dg - \bbi a_2^\dg) = e_2 a_-^\dg \equiv a_{-2}^\dg
\lbl{6.17}
\ee
Replacing in Eqs\,(\ref{6.16}),(\ref{6.17}) $a$ with $b$, we obtain the basis vectors in momentum
representation, considered in Eq.\,(\ref{6.1}) and in Table 1. The weak isospin $I_3$ is equal
to the eigenvalues of the operator $e_1/2 = \sg_3/2$, acting on the states created by
$b_{+1}^\dg$, $b_{+2}^\dg$, $b_{-2}^\dg$, $b_{-1}^\dg$, 
The pair $(b_{-2}^\dg,b_{-1}^\dg)$ is associated with the weak isospin doublet $(\nu_e, e^-)$,
whilst the pair$(b_{+1}^\dg,b_{+2}^\dg)$ is associated with the doublet of the new particles
$(\epsilon^+,\nu_\epsilon)$, predicted by this model.
 
The electric charge of each of those particles is given by the usual formula
$Q= \frac{y}{2} + I_3$. The hypercharge operator is equal to the generator ${\hat p}_5 = -i \p/\p x^5$
of the translations along the fifth dimension, as explained in Sec.\,3.3. The occurrence of
electric charge due to the 5th dimension was considered in Sec.\,3.3 as an alternative to the occurrence of electric charge
as being due to the doubling of the Dirac field \`a la Eq.\,(\ref{3.42}). In this section, we
have shown that if we also consider weak interaction which requires additional doubling of the
fields, the electric charge is a result of both contributions. This is in agreement with the results
of Refs.\ci{Cotaescu,PavsicKaluzaLong} where it was found that within the framework of a higher
dimensional theory \`a la Kaluza-Klein, a gauge charge is the sum of the orbital angular momentum
in the extra, so-called internal, space, and the spin momentum in that extra space. Therefore, also
the weak interaction charge is predicted to contain two such contributions, and so besides the term $g I_a$, $a=1,2,3$,
coupled to the gauge field ${W_\mu}^a$, there must also be a term of the form  ${k_a}^{\bar M} \p_{\bar M}$,
where  ${k_a}^{\bar M}$ are Killing vector fields and $\p_{\bar M}\equiv \p/\p x^{\bar M}$ is the partial derivative
with respect to the coordinates $X^{\bar M}$ of the internal space. From this point of view, the usual
electroweak theory  and chromodynamics should be completed to take into account such extra term as well.

\subsection{The stability in the presence of negative energies}

The Dirac wave function contains negative energy components. This is considered as problematic,
and it is usually stated that the problem of negative energies is resolved within the framework
of quantum field theory. But, in view of the
findings in Refs.\,\ci{Smilga1,Smilga2,Smilga3,Smilga4,PavsicStable,Kaparulin,PavsicPUReview,Deffayet}, the presence
of negative energy states does not pose any problems in physically realistic situations. This
also is true for the wave function, satisfying the Dirac equation. Namely, in a realistic
situation in which the potential $A_\mu$ is bounded from below and from above, the solutions
of the Dirac equation cannot evolve toward infinity. Its real and imaginary components remain
finite. Besides that, from the very construction of the theory, we have the conserved current $j^\mu =
{\bar \psi} \gam^\mu \psi$. Therefore, $\int \dd^3 \bx \, \psi^\dg \psi$ remains constant and finite in time, and so does $\int \dd^3 \bx \, \psi^\dg \al^i \psi$. The expectation value of
the gauge-invariant
(kinetic) momentum ${\hat \pi}_\mu = - i \DD_\mu = - i (\p_\mu - \bbi e A_\mu)$ is
$\langle {\hat \pi}_\mu \rangle = \int \dd^3 \psi^\dg{\hat \pi}_\mu \psi$, and it also does not escape
into infinity. This can be seen by calculating the time derivative
$$
  \frac{\dd}{\dd t}\langle {\hat \pi}_\mu \rangle = \int \dd^3 \bx 
  	\left ( {\dot \psi}^\dg (-i {\DD}_\mu )\psi
  	 + \psi^\dg {\DD}_\mu) {\dot \psi} + \psi^\dg \frac{\p {\hat \pi}_\mu}{\p t} \psi
  	 \right )  \hs{3cm}$$
$$ = -i \int \dd^3 \bx \psi^\dg \left ( \left [-i \DD_\mu\,,-\bbi e A_0 + \beta m - i \al^i \DD_i \right ] 
    +  \frac{\p {\hat \pi}_\mu}{\p t} \right ) \psi$$
\be
   = \int \dd^3 \bx\, \psi^\dg \gam^0 (-i \bbi) e F_{\mu \nu} \gam^\nu \psi~,
  ~~~~~~~~~~~~~~~-i \bbi = e_1 = \sg_3 .
\lbl{6.2B}
\ee
This is the expectation value of the operator equation
\be
  \frac{\dd {\hat \pi}_\mu}{\dd t} = \frac{\p {\hat \pi}_\mu}{\p t} + i [H,{\hat \pi_\mu}]
    = -i \bbi e F_{\mu \nu} \gam^\nu  ,
\lbl{6.2aB}
\ee
 the operator equivalent of the Lorentz force equation. In the literature the quation (\ref{6.2aB})
 is considered to have a limited validity, because of the presence of negative energies which are
 believed to lead to paradoxes.
 
 However, in view of the finding that negative energies are not problematic at all, the equations
(\ref{6.2B}) and (\ref{6.2aB}) are physically viable as well. Namely, because 
$\int \dd^3 \bx {\bar \psi} \gam^\mu \psi$ remains finite during the evolution of the system, also
$\int \dd^3 \bx {\bar \psi} F_{\mu \nu} \gam^\mu \psi$ and thus
$\dd \langle {\hat \pi}_\mu \rangle/\dd t$ remain finite in the presence of a properly
bounded $F_{\mu \nu}$. A system described by the Dirac equation does not have instabilities due to
the presence of negative energies, provided that $A_\mu$, and consequently 
$F_{\mu \nu} = \p_\mu A_\nu - \p_\nu A_\mu$, is bounded from below and from above.

A free electron cannot emit a photon, because this would violate 4-momentum conservation. This is
true regardless of whether the electron has positive or negative energy. The electron in
interaction (e.g., in an atom or in whatever potential) can radiate a photon. The electron, whose
energy is positive, can jump from a higher to a lower energy level in a potential bounded from below.
The electron, whose energy is negative, is not bound in such potential bounded from below.
It rolls upwards in such potential, and if the potential is bounded also from above,
then it escapes from such regions and thereafter moves as a free electron with a constant velocity.
The fact that
negative kinetic energies are available for the electron does not imply its rolling
down to lower and lower energies. A free electron cannot roll at all, and how the electron behaves in
the presence of potential depends on the shape
of the potential\ci{Choi1,Choi2,Choi3,PavsicStable,PavsicPUReview,PavsicStumbling}. It remains to analyze
the situation within the framework of quantum field theory. In view of Ref.\ci{PavsicFirence,PavsicPUReview}, where the notorious
vacuum instability, presumably causing an instantaneous vacuum decay, was reconsidered afresh and found that no
such problems occur, it is reasonable to expect that a QFT description of electrons interacting with photons makes
sense even if one allows for negative energy states.

\subsection{The wave packet localization and Lorentz covariance}

 A wave function $\psi(t,\bx) = \sqr \left ( \sqrt{\om_\bx} \vphi(t,\bx)
  + \frac{i}{\sqrt{\om_\bx}} {\dot\vphi} (t,\bx) \right )$ is defined with respect to a particular
  spacetime split $(1+3)$ and determines the probability density of observing at time $t$ the particle
  at position $\bx$. This means that $\psi(t,\bx)$ is defined with respect to an inertial (Lorentz) system
  $S$ in which an observer measures the 3D position of the particle described by $\psi(t,\bx)$. A state
  is given by a vector
  \be
    |\Psi \rangle = \psi^{(\bx)} a_{(\bx)}^\dg \vac \equiv
    \int \dd^3 \bx \, \psi(t,\bx) a^\dg (\bx) \vac ,
\lbl{6.3A}
\ee
where $a^\dg (\bx)$ creates a particle at $\bx$. 

The 3D position is thus determined on a given 3-surface $\Sigma$ which in the above particular case
is just $t=0$. But in general, it can be any surface. To embrace such general case, let us write the
state in the following covariant form\ci{PavsicRelatWaveFun2}:
\be
   |\Psi \rangle = \int \dd \Sigma \, \psi(s_\Sg,{\bar x}_\Sg) a^\dg (\barx_\Sigma) \vac ,
\lbl{6.3aA}
\ee
where $s_\Sg$ is the proper time along a worldline orthogonal to a 3-surface $\Sg$, and 
$\barx_\Sg \equiv \barx_\Sg^\mu = ({\delta^\mu}_\nu - n_\Sg^\mu n_{\Sg \nu}) x^\nu$ are the
projections of the spacetime coordinates onto $\Sg$, i.e., $\barx_\Sg \equiv \barx_\Sg^\mu$
are coordinates within $\Sg$.

A state $|\Psi \rangle$ is invariant under Lorentz transformations. In another inertial system $S'$,
the same state $|\Psi \rangle$ is given by the expression
\be
|\Psi \rangle = \int \dd \Sigma \, \psi'(s'_\Sg,{\bar x}'_\Sg) a'^\dg (\barx'_\Sigma) \vac ,
\lbl{6.4A}
\ee
where the integration is over the same 3-surface $\Sg$. The state $|\Psi \rangle$ in Eq.\,(\ref{6.4A})
is the same as in Eq.\,(\ref{6.3aA}), only the components $\psi'(s'_\Sg,{\bar x}'_\Sg)$ and
the basis vectors $a'^\dg (\barx'_\Sigma)$ are different functions of the new coordinates.
Namely, under a Lorentz transformation $\barx' = L \barx$, we have\footnote{
A covariant expression for the wave function is (see Ref.\ci{PavsicRelatWaveFun2})\\	
$\psi(s,\barx) = \sqr \left ( \sqrt{\om_\barx} \vphi (s,\barx) + \frac{i}{\sqrt{\om_\barx}} 
\frac{\dd \psi}{\dd s}(s,\barx)\right )$, where $\om_\barx = \sqrt{m^2 + \eta^{\mu \nu} {\bar \p}_\mu {\bar \p}_\nu}$, and similarly for $a^\dg(s,\barx)$ and $a(s,\barx)$. Because
$\om_\barx = \sqrt{m^2 + \eta^{\mu \nu} {\bar \p}_\mu {\bar \p}_\nu}$
$=\sqrt{m^2 + \eta^{\mu \nu} {\bar \p}'_\mu {\bar \p}'_\nu}$ $=\om_{\barx'}$, and because
$\vphi(s,\barx)= \vphi'(s',\barx')$, $\dd \vphi (s,\barx)/\dd s=$ $\dd \vphi' (s',\barx')/\dd s'$,
we have $\psi(s,\barx) = \psi'(s',\barx')$ and $a(s,\barx) = a' (s',\barx')$.
}
$s' = s$,  
$a^\dg (\barx_\Sigma) = a^\dg(L^{-1} \barx'_\Sg) = a'^\dg (\barx'_\Sg)$ and
$\psi(s_\Sg,{\bar x}_\Sg) = \psi(s_\Sg,L^{-1}{\bar x}_\Sg)$ $=\psi'(s'_\Sg,{\bar x}'_\Sg)$.
In both cases, (\ref{6.3aA}) and (\ref{6.4A}), the operators $a^\dg(\barx_\Sg)$ and  $a'^\dg(\barx'_\Sg)$
create particles on the same 3-surface $\Sg$, only the reference frame (the inertial system
and the associated coordinate systems) are different.

In the above discussion, we considered a {\it passive Lorentz transformation}, i.e., how a state 
$|\Psi \rangle$ is expressed in different inertial (Lorentz) systems $S$ and $S'$\footnote{
	An observer in the station can write the wave function for an electron in the double-slit
	experiment performed in the train. It differs from the wave function of the same electron
	as written (determined) by the observer in the train. Both wave functions represent the same state.}.
We can also consider an {\it active Lorentz transformation} between the different states $|\Psi \rangle$
and $|\Psi' \rangle$, as observed in a given (fixed) inertial system\footnote{
	An observer in the station examines two equivalent double slit experiments: one in the station,
	and another one in the train. He or she finds that the wave function of the electron in one experiment
	differs from the wave function of the electron in the other experiment. Those two wave functions
	represent two different states.}. One such state is given in Eq.\,(\ref{6.3aA}) and the other one is
\be
  |\Psi'\rangle = \int \dd \Sigma \psi' (s_\Sigma,{\bar x}_\Sg) a^\dg (\barx_\Sg) \vac,
\lbl{6.5A}
\ee
where $\barx_\Sg$ are the coordinates of the same surface $\Sg$, while the wave function is different.

For a better understanding of the concept of a state expressed in terms of a wave function and
position states on a given 3-surface, let us further examine the situation of a double-slit experiment on the train and in
the station (see footnotes 10 and 11).

The double-slit experiment can be performed: (i) with the equipment on the train, or (ii) with the
equipment in the station. The state of the electron in the case (i), $|\Psi_T\rangle$, is different
from the state of the electron in the case (ii), $|\Psi_S \rangle$.

For the case (i), we have
\be
  |\Psi_T\rangle = \int \dd \Sg_T \, \psi_T (s_T,\barx_T) a^\dg (\barx_T) \vac
  =  \int \dd \Sg_S \, \psi''_T (s_S,\barx_S) a^\dg (\barx_S) \vac ,
\lbl{6.7A}
\ee
where $\psi_T (s_T,\barx_T)$ is the electron wave function of the experiment on the train, expressed
in terms of position $\barx_T \equiv \barx_T^\mu$ on the train. Here $a^\dg (\barx_T)$ are the operators
that create particles at positions $\barx_T^\mu$ which are all simultaneous on the train (i.e.,
$\barx_T^\mu$ denote points on the train's simultaneity surface $\Sg$). Thus, the state $|\Psi_T \rangle$ is
expanded in terms of the basis vectors $a^\dg (\barx_T) \vac \equiv |\barx_T \rangle$, which are position
states on the train.  Alternatively, the same state $|\Psi_T \rangle$ can also be expanded in terms
of the basis vectors $a^\dg (\barx_S) \vac \equiv |\barx_S \rangle$, which are position states
in the station, and so $\barx_S \equiv \barx_S^\mu$ are simultaneous in the station.
 
For the case (ii), we have
\be
|\Psi_S\rangle = \int \dd \Sg_S \, \psi_S (s_S,\barx_S) a^\dg (\barx_S) \vac
=  \int \dd \Sg_T \, \psi'''_S (s_T,\barx_T) a^\dg (\barx_T) \vac .
\lbl{6.8A}
\ee
In distinction to Eq.\,(\ref{6.7A}), the expansion coefficients $\psi'''_S (s_T,\barx_T)$ (i.e., the wave
function) in the $a^\dg (\barx_T)\vac \equiv |\barx_T \rangle$ basis are different from the expansion
coefficients $\psi_T (s_T,\barx_T)$ in the same basis $a^\dg (\barx_T)\vac \equiv |\barx_T \rangle$.

Thus $\psi_T (s_T,\barx_T)$ and $\psi'''_S(s_T,\barx_T)$ represent, respectively, two different
states $|\Psi_T \rangle$ and $|\Psi_S \rangle$, expanded in the train position states
$|\barx_T \rangle \equiv a^\dg (\barx_T) \vac$.

Similarly, $\psi''_T(s_S,\barx_S)$ and $\psi_S (s_S,\barx_S)$ represent, respectively, the states
$|\Psi_T \rangle$ and $|\Psi_S \rangle$, expanded in terms of the station position states
$|\barx_S \rangle \equiv a^\dg (\barx_S) \vac$.

The position on the train can be expressed in terms of any coordinates (see Eqs.\,\ref{6.3aA}) and (\ref{6.4A})).
It can be expressed in term sof the coordinates ${\barx_T}^{\,\mu}$ of the inertial system $S$, or it can be expressed
in terms of the coordinates $\barx'_T$$^{\mu}$ of some other inertial system $S'$. The same holds
for the coordinates ${\barx_S}^{\,\mu}$ of the position in the station.

To sum up, a given state $|\Psi \rangle$ can be expressed in terms of position states of any 3-surface,
e.g., $\Sg_T$ or $\Sg_S$ (see Eqs.\,(\ref{6.7A}) and (\ref{6.8A})) or any other 3-surface. The same
state $|\Psi \rangle$ can be described in terms of coordinates of any inertaial system, e.g., $S$ or $S'$
(see Eqs.\,(\ref{6.3aA}) and (\ref{6.4A})). One thing is how a state $|\Psi \rangle$ is expanded, and
the other thing is the inertial system from which $|\Psi \rangle$ is observed.

The position of the electron in the double-slit experiment on the train is measured, e.g., with a
fluorescent screen. That screen, which is at rest with respect to the train can be observed from the
station. Hence, the position states $a^\dg ({\barx_T}) \vac \equiv |\barx_T \rangle$, in particular the
position states on the screen on the train, can as well be described from the point of view of the station
(reference frame $S'$). So we have $|\barx_T \rangle$ = $a^\dg ({\barx_T}) \vac =a'^\dg (\barx'_T) \vac$
$=|\barx'_T \rangle$. The corresponding expansion coefficients (wave function) are $\psi_T (\barx_T)$
and $\psi'_T (\barx'_T)$, respectively. They describe the same state $|\Psi_T \rangle$, expressed in terms
of the train position states (i.e., measured with the screen on the train).
Analogous holds for the position of the electron in the double-slit experiment in the station.

Wave packets move at speeds slower or equal to the speed of light. This means that the expectation
value of position lies on a worldline within the light cone. Because the wave packets are spread, they
have tails that traverse the light cone. This fact has been usually interpreted as a violation of
causality\ci{Hegerfeldt,Hegerfeldt1,Hegerfeldt2,Rosenstein,Mosley,Eckstein,Barat}
and thus a strong argument against the relativistic wave function as being a physically
viable concept. A different view is maintained in Refs.\ci{Valente,Karpov,Antoniou,Fleming2,Wagner,Ruijsenaars,PavsicRelatWaveFun1,PavsicStumbling} in which relativistic wave function
does not violate causality in the sense of being able to transmit information faster than light.

\section{Conclusion}

In view of the findings\ci{Smilga1,Smilga2,Smilga3,Smilga4,PavsicFirence,PavsicStable,PavsicPUReview,Kaparulin,PavsicStumbling,Deffayet} according to which the presence of negative energies does not
necessarily leads to instabilities, we have revisited relativistic quantum mechanics and quantum
field theory. So the Dirac wave function can be interpreted as the probability amplitude without the usual restrictions, though it contains negative energies. Creation and annihilation operators,
namely, bosonic and fermionic quantum fields, are the phase space basis vectors in the Witt basis. 
They generate infinite-dimensional symplectic and orthogonal Clifford algebras, respectively. One of its
elements is the bare vacuum and another one is the Dirac vacuum. The electric charge arises on the one hand
from doubling the Dirac fields, which involves an additional imaginary unit $\bbi$, and on the other, hand
it arises from the fifth dimension. By considering the $\bbi$ as one of the three quaternionic
imaginary units, the Dirac
field was extended to have values in the quaternionic algebra which can be rewritten as the complexified Clifford algebra $Cl(2)\otimes {\mathbb C}$. It contains the weak isospin doublet $(\nu_e,e^-)$ and also
an additional doublet $(\nu_\epsilon,\epsilon^+)$ that forms the representation of another copy of $SU(2)$.
So the theory predicts new leptons that together with the corresponding new quarks $(u',d')$ are expected
to be invisible to ordinary particles, and thus form dark matter.

\end{document}